\documentclass[a4paper,fleqn]{cas-dc}

\usepackage[authoryear,longnamesfirst]{natbib}
\usepackage{hyperref}

\usepackage{amsmath}
\usepackage{amsfonts} 
\usepackage{amssymb}
\usepackage{amsthm}

\usepackage{algorithmic}
\usepackage{algorithm}

\usepackage{array}
\usepackage{booktabs}
\usepackage{multirow}

\usepackage{graphicx}
\usepackage[caption=false,font=normalsize,labelfont=sf,textfont=sf]{subfig}
\usepackage{tikz}
\usetikzlibrary{positioning}

\usepackage{textcomp}
\usepackage{url}
\usepackage{verbatim}
\usepackage{xcolor}
\usepackage{xspace}
\usepackage{setspace}
\usepackage{enumitem}

\usepackage{stfloats}

\usepackage{listings}

\usepackage{tcolorbox}

\definecolor{dkgreen}{rgb}{0,0.6,0}
\definecolor{dkblue}{rgb}{0,0,0.8}
\definecolor{mauve}{rgb}{0.58,0,0.82}
\definecolor{todocolor}{rgb}{0.9,0.1,0.1}

\makeatletter
\newcommand\runinhead{\@startsection{paragraph}{4}{\z@}%
   {-6\p@}%
   {-6\p@}%
   {\normalfont\normalsize\bfseries\boldmath
    \rightskip=\z@ \@plus 8em\pretolerance=10000 }}
\makeatother

\newcommand{\tool}{\textsc{Cg-FsPta}\xspace}
\newcommand{\wave}{\textsc{Wave}\xspace}
\newcommand{\sfr}{\textsc{Sfr}\xspace}

\newcommand{\ruledef}[3]{ #1 $\frac{\begin{array}[c]{c}#2\end{array}}{\begin{array}[c]{c}#3\end{array}}$}
\newcommand{\rulename}[1]{\rulelab{\MakeUppercase{#1}}}
\newcommand{\rulelab}[1]{\small\texttt{[{#1}]}}

\newcommand{\addr}[3]{\scriptsize{#1\!\xrightarrow{\texttt{Addr}}\!#2}}
\newcommand{\cpy}[3]{\scriptsize{#1\!\xrightarrow{\texttt{Copy}}\!#2}}
\newcommand{\cp}[2]{\scriptsize{#1\!\xrightarrow{\texttt{Copy}}\!#2}}
\newcommand{\gep}[3]{\scriptsize{#1\!\xrightarrow{\texttt{Gep}.#3}\!#2}}
\newcommand{\load}[3]{\scriptsize{\!#1\xrightarrow{\texttt{Load},~#3}\!#2}}
\newcommand{\store}[3]{\scriptsize{\!#1\xrightarrow{\texttt{Store},~#3}\!#2}}

\newcommand{\lab}{\ell\xspace}

\newcommand{\calV}{\mathcal{V}\xspace}

\newcommand{\calP}{\mathcal{P}\xspace}
\newcommand{\calO}{\mathcal{O}\xspace}

\newcommand{\pts}{\textit{pts}}

\newcommand{\cupeq}{~\cup\!=}

\theoremstyle{definition}

\newtheorem{example}{Example}

\newcommand{\code}[1]{{\fontfamily{cmtt}\fontseries{m}\fontshape{n}\selectfont\small{#1}}}

\let\oldnl\nl
\newcommand{\nonl}{\renewcommand{\nl}{\let\nl\oldnl}}

\newcommand{\stimes}{\texttimes\xspace}

\newtcolorbox{answerBox}[1][]{
   colback=gray!10,
   colframe=gray!10,
   boxrule=0.2mm,
   arc=0.5mm,
   width=\columnwidth,
   left=0.2mm,
   right=0.2mm,
   top=0.2mm,
   bottom=0.2mm,
   #1
}

\begin{document}

\let\WriteBookmarks\relax
\def\floatpagepagefraction{1}
\def\textpagefraction{.001}

\shorttitle{Flow-Sensitive Pointer Analysis without CFG}    

\shortauthors{J. Zhang et al.}

\title[mode = title]{Flow Sensitivity without Control Flow Graph: An Efficient Andersen-Style Flow-Sensitive Pointer Analysis}

\author[1]{Jiahao Zhang}[
      ]
\ead{jiahao.zhang6@unswalumni.com}
\fnref{eq}

\author[1]{Xiao Cheng}[
      ]
\ead{xiao.cheng@unsw.edu.au}
\fnref{eq}

\author[1]{Yuxiang Lei}[
      ]
\ead{youthyoungray@gmail.com}
\corref{cor1}

\affiliation[1]{organization={School of Computer Science and Engineering, University of New South Wales},
            city={Sydney},
            country={Australia}}


\cortext[cor1]{Corresponding author}

\fntext[eq]{These authors contributed equally to this work.}

\begin{abstract}
Flow-sensitive pointer analysis constitutes an essential component of precise program analysis for accurately modeling pointer behaviors by incorporating control flows. Flow-sensitive pointer analysis is extensively used in alias analysis, taint analysis, program understanding, compiler optimization, etc.
Existing flow-sensitive pointer analysis approaches, which are conducted based on control flow graphs, have significantly advanced the precision of pointer analysis via sophisticated techniques to leverage control flow information.
However, they inevitably suffer from computational inefficiencies when resolving points-to information due to the inherent complex structures of control flow graphs.

We present \tool, a \textit{Flow-Sensitive Constraint Graph} (FSConsG) based flow-sensitive pointer analysis to overcome the inefficiency of control-flow-graph-based analysis.
\tool uses a flow-sensitive variant to leverage the structural advantages of set-constraint graphs (which are commonly used in flow-insensitive pointer analysis) while keeping the flow sensitivity of variable definitions and uses, allowing the incorporation of sophisticated graph optimization and dynamic solving techniques.
In this way, \tool achieves significant efficiency improvements while keeping the precision of flow-sensitive analysis.
Experimental evaluations on benchmark programs demonstrate that \tool, which leverages the FSConsG to simplify graph structure and significantly reduces both memory usage and execution time while maintaining precision.
In particular, by solving in the FSConsG, \tool achieves an average memory reduction of 33.05\% and accelerates flow-sensitive pointer analysis by 7.27\stimes compared to the state-of-art flow-sensitive pointer analysis method. These experimental results underscore the efficacy of \tool as a scalable solution to analyze large-scale software systems, thus establishing a robust foundation for future advancements in efficient program analysis frameworks.

\end{abstract}


\maketitle

\section{Introduction}
\label{sec:introduction}

Pointer analysis determines the potential memory locations a pointer may reference during runtime, serving as a foundational component of static program analysis. This analysis enables numerous applications across software security~\cite{sui_detecting_2014,fink2008effective,yan_spatio_2018,shi2021fusion,yulei2012saber,machiry2017dr,li_path_2022,guo2024precise}, program verification~\cite{lee2005automatic, das2002esp, xiao2024fgs}, and compiler optimization~\cite{ovs, hardekopf2007exploiting}. Enhancing the efficiency and precision of pointer analysis~\cite{andersen1994, hardekopf_ant_2007, pearce_efficient_2007, sui_detecting_2014,tan2021making,zuo2021chianina} represents a long-standing research challenge, as its effectiveness directly influences the quality of downstream static analysis tasks.

The performance of pointer analysis depends significantly on the analysis approach and the \emph{underlying graph representation}. Flow-insensitive pointer analysis, which disregards control flow and execution order, achieves exceptional performance due to its simplicity and reduced computational demands~\cite{hardekopf_ant_2007, pereira_wave_2009}. 
It typically operates on a \emph{constraint graph (ConsG)}, where each node represents a single variable, and each edge encodes a constraint between two variables. This representation is efficiently solved using Andersen-style algorithms~\cite{andersen1994, hardekopf_ant_2007}, which have been the subject of extensive optimization research for decades~\cite{pereira_wave_2009,chang_fast_2019,liu2022pus}.
However, this computational efficiency comes at the cost of precision, as flow-insensitive analysis produces overly conservative points-to sets by aggregating all possible program states without considering execution order, frequently resulting in spurious points-to relationships~\cite{yu_level_2010, sui_detecting_2014}. These imprecisions, inherent to the lack of flow sensitivity, render such analyses inadequate for applications requiring detailed modeling of program execution sequences. Conversely, flow-sensitive pointer analysis provides substantially higher precision by accounting for the sequential execution order of program statements. 
By tracking value changes at specific program points, flow-sensitive pointer analysis enables precise updates of points-to information.
This enhanced precision makes flow-sensitive analysis particularly valuable in modern program analysis scenarios, especially for tasks where correctness and detailed behavioral modeling are critical~\cite{sui_sparse_2016, hardekopf_flow-sensitive_2011}. 

In general, flow-sensitive pointer analysis is conducted based on a \emph{control flow graph (CFG)}, techniques including semi-sparse approach~\cite{hardekopf_semi-sparse_2009}, level-by-level analysis~\cite{yu_level_2010}, and the sparse value-flow graph (SVFG)~\cite{sui_sparse_2016}.
These approaches enhance pointer analysis precision by selectively propagating points-to information or carefully partitioning the program into segments. Although semi-sparse techniques~\cite{hardekopf_semi-sparse_2009} reduce redundant computations through data-dependency management and selective updates, they frequently suffer from scalability issues. 
Similarly, the SVFG achieves flow-sensitivity by maintaining sparse def-use chains generated from the integration of a CFG and the result of a flow-insensitive pointer analysis, which is essentially a simplified CFG. Consequently, solving the SVFG involves both processing statements in the nodes and propagating values along the edges.
The overlapped variables in adjacent nodes lead to repeated computation and thus cause inefficiency.
Furthermore, since the graphs of the current flow-sensitive pointer analysis techniques bundle program statements into the nodes, existing state-of-the-art graph simplification~\cite{hardekopf_ant_2007,ovs,lei2023recursive} and dynamic solving techniques~\cite{pereira_wave_2009,liu2022pus} are not applicable.

This observation motivates a fundamental research question: \emph{can we synthesize the precision of flow-sensitive pointer analysis with the computational efficiency of Andersen-style flow-insensitive analysis?} 
To address this challenge, we introduce \tool, a novel flow-sensitive pointer analysis approach based on the \textit{Flow-Sensitive Constraint Graph (FSConsG)}, which effectively bridges the gap between efficient Andersen-style constraint-based analysis and precise flow-sensitive pointer analysis.
Unlike existing CFG-based techniques such as SVFG \cite{sui_sparse_2016}, where nodes encapsulate statements potentially containing multiple variables, thereby preventing the application of powerful optimization techniques developed for Andersen-style pointer analysis, our FSConsG maintains the elegant simplicity of ConsG structures: each node represents exactly one variable, with no embedded control flow information.
This critical insight—separating variable representation from control flow—yields a significantly cleaner and more computationally efficient graph representation compared to traditional flow-sensitive approaches. 
Our FSConsG is inherently compatible with the state-of-the-art graph optimization techniques, including variable subsitution~\cite{ovs} and cycle elimination \cite{tarjan1972depth}, as well as efficient constraint solvers like \wave \cite{pereira_wave_2009} and \sfr \cite{lei2019fast}.
Consequently, \tool substantially reduces computational overhead while maintaining the precision characteristics of flow-sensitive analysis, demonstrating its potential as a scalable and effective solution for current program analysis challenges.

The construction of an FSConsG requires the information of a CFG and the result of a flow-insensitive pointer analysis, 
similar to the SVFG used in existing techniques VSFS~\cite{barbar_object_2021,hardekopf_semi-sparse_2009,sui_demand-driven_2017}.
The essential difference is that the graph we used is constructed in the form of ConsG, i.e., each variable is kept as a node, and each edge represents an instruction. 
In particular, in an FSConsG, we separate address-taken variables (abstract memory objects) into versions according to the execution points (CFG nodes) and use def-use relations (calculated by the approach proposed in \cite{hardekopf_semi-sparse_2009}) to construct a series of def-use chains of the address-taken variables, maintaining the flow-sensitivity of address-taken variables.
In this form, we are able to maintain a points-to set for each node and solve points-to relations by processing only the edges.
Since our pointer analysis is conducted on LLVM's intermediate representation (LLVM-IR), where top-level variables (pointers) are already field-sensitive (in an SSA form), our FSConsG is sufficient to perform flow-sensitive pointer analysis.
Based on the FSConsG, we propose a flow-sensitive constraint solver.
With specific mechanisms for solving address-taken variables, the solver guarantees its soundness and precision by outputting points-to-set results identical to that of existing techniques, e.g., SFS and the semi-sparse approach \cite{hardekopf_semi-sparse_2009}.

Our \tool integrates flow-insensitive and flow-sensitive pointer analyses into the same abstract program model, simplifying the graph representation of flow-sensitive pointer analysis, and uniquely boosting the speed of flow-sensitive pointer analysis by minimizing computational overhead, with the incorporation of powerful graph simplification and dynamic solving techniques.

We implemented our \tool and evaluated its scalability and efficiency by comparing it to the state-of-the-art VSFS.
Integrated into the LLVM-16 compiler infrastructure and evaluated using the SPEC CPU 2017 benchmark suite, \tool consistently outperforms VSFS in the number of constraints within the graph, memory usage, and execution time.  
These improvements stem from effectively reducing the number of nodes and edges in the FSConsG without sacrificing analytical precision. Moreover, the constraint-solving algorithm optimizes memory usage and accelerates execution, enabling \tool to achieve significantly faster average execution compared to VSFS.

Our key contributions are as follows:

\begin{itemize}
    \item We integrate flow-insensitive and flow-sensitive pointer analyses into the same program model via flow-sensitive constraint graph (FSConsG), reducing the graph complexity of flow-sensitive pointer analyses and making flow-sensitive pointer analyses well compatible with the state-of-the-art graph simplification and constraint solving techniques.

    \item Based on the FSConsG, we propose a constraint-solving algorithm with specific mechanisms to handle address-taken variables. By incorporating with the state-of-the-art techniques, the solver significant boosts the efficiency of flow-sensitive pointer analysis.

    \item We integrated \tool --- including both the construction of FSConsGs and the constraint solver --- into the LLVM-16 infrastructure and conducted a comprehensive evaluation based on benchmark programs. The experimental results indicate that, compared to VSFS, \tool significantly enhances efficiency by reducing memory consumption, achieving an average reduction of 33.05\%, and demonstrating an average execution time that is 7.27$\times$ faster. These results highlight its effectiveness and scalability in analyzing modern C/C++ programs.
    
\end{itemize}

The remainder of this paper is organized as follows: Section 2 provides the necessary background on pointer analysis, detailing the methodologies of set constraint-based analysis, flow-insensitive analysis, and flow-sensitive analysis. Section 3 explains the motivation for developing the \tool, addressing the limitations of existing methods. Sections 4 and 5 present our approach, describing the construction of FSConsG and the solver of \tool, as well as the application of graph optimization techniques such as graph folding. Section 6 evaluates our approach using experiments, comparing \tool with the state-of-the-art method in terms of graph structure,  memory usage, and execution time. Section 7 discusses related work, situating our contributions in the context of prior research. Finally, Section 8 concludes with a summary of findings and potential directions for future work.

\section{Background and Problem Formulation}
\label{sec:background}

This section presents the foundational concepts underpinning our approach. We begin by describing the target language and introducing set-constraint based pointer analysis. We then compare flow-insensitive (constraint graph-based) and flow-sensitive (control-flow graph-based) pointer analysis methodologies, highlighting their respective advantages and limitations. Finally, we formulate the central problem addressed in this work: how to leverage the computational efficiency of Andersen's pointer analysis to resolve flow-sensitive analysis challenges without sacrificing the precision benefits inherent to flow-sensitivity.

\subsection{Analysis Domains, LLVM Instructions and Constraints}
\label{sec:LLVMIR}

\begin{table*}[t]
    \centering
    \caption{\label{tab:domain} Analysis domains, LLVM instructions, and constraint edges.}
    \addtolength{\tabcolsep}{-.3ex}
    \begin{tabular}{l|l}
        \hline
    \scalebox{1.0}[1]{
    \small
    \renewcommand{\arraystretch}{1.0}
    \begin{tabular}[t]{l@{\hspace{1mm}}l@{\hspace{1mm}}l@{\hspace{2mm}}l}
    \multicolumn{4}{c}{Analysis Domains}\\
    \hline
    \addlinespace[1pt]
    $\lab$ & $\in$ & $\mathcal{L}$ & Statements\\
    $i,j,w$ & $\in$ & $\mathcal{C}$ & Integer constants\\
    $s$ & $\in$&$\mathcal{S}$ & Stack virtual registers\\
    $g$ & $\in$&$\mathcal{G}$ & Global variables\\
    $f$ & $\in$&$\mathcal{F}\subseteq \mathcal{G}$ & Program functions\\
    $p, q, r,x,y$ & $\in$ & $\calP=\mathcal{S}\cup\mathcal{G}$ & Top-level variables\\
    $o,a,b,c,o.f_i$ & $\in$ & $\calO$ & Address-taken variables\\
    $u,v$ & $\in$ & $\calV\!=\!\calO \cup \calP$ & Program variables\\
    \end{tabular}
    }
    &
    \scalebox{1.0}[1]{
    \small
    \renewcommand{\arraystretch}{1.15}
    \begin{tabular}[t]{l@{\hspace{2mm}}l@{\hspace{3mm}}l}
    Instruction & Constraint & Type\\
    \hline
    \addlinespace[3pt]
    $\mathtt{p=\&o}$ & $p\xleftarrow{\textsf{Addr}}o$ & \textsf{Addr}\\
    $\mathtt{p=q}$ & $p\xleftarrow{\textsf{Copy}}q$ & \textsf{Copy} \\
    $\mathtt{p=\&q\!\rightarrow\! fld}$  & $p\xleftarrow{\textsf{Gep, fld}}q$ & \textsf{Gep}\\
    $\mathtt{p=*q}$ & $p\xleftarrow{\textsf{Load}}q$ & \textsf{Load}\\
    $\mathtt{*p=q}$ & $p\xleftarrow{\textsf{Store}}q$ & \textsf{Store}\\
    \end{tabular}
    }
    \\
    \hline
    \end{tabular}
    \vspace{-0mm}
\end{table*}

Following previous works \cite{hardekopf_ant_2007,sui_-demand_2016,chang_fast_2019}, we conduct our pointer analysis on an LLVM-like intermediate representation \cite{lattner2004llvm}.
Table~\ref{tab:domain} presents the analysis domains and LLVM instructions relevant to flow-sensitive pointer analysis.
This intermediate representation employs a partial static single assignment (SSA) form, wherein the set of all program variables $\calV$ is partitioned into two disjoint subsets:
\emph{address-taken variables} $\calO$, comprising all abstract memory objects and their fields, and
\emph{top-level variables} $\calP$, consisting of stack virtual registers ($\mathcal{S}$, denoted by symbols prefixed with \textsf{``\%"}) and global variables ($\mathcal{G}$, denoted by symbols prefixed with \textsf{``@"}). 
Top-level variables are represented in explicit SSA form, allowing direct access and manipulation.
In contrast, address-taken variables in $\calO$ are maintained in non-SSA form and can only be accessed indirectly through top-level variables via \textsf{Load} or \textsf{Store} instructions.

A key consequence of this design is that there are no direct assignments between address-taken variables in LLVM-IR. Any source-level assignment to an address-taken variable is compiled into a \texttt{store} instruction through a top-level pointer, and any read from an address-taken variable is compiled into a \texttt{load} instruction. Furthermore, multi-level pointer operations in the source language (e.g., \texttt{**p}) are decomposed by the compiler into sequences of single-level \texttt{load}/\texttt{store} instructions, each operating through a top-level temporary in SSA form. This two-level design ensures that every pointer dereference in LLVM-IR involves exactly one level of indirection.

The C-like representation of LLVM instructions and their corresponding constraint edges are presented in Table~\ref{tab:domain} (right side). 
The \textsf{Addr} constraint establishes the fundamental mapping between pointers and their referenced memory objects, initializing the basis for all pointer relationships. The \textsf{Copy} constraint represents direct assignment between top-level variables, enabling value propagation through the program. For memory access operations, \textsf{Load} and \textsf{Store} constraints capture the bidirectional semantics of retrieving values from and writing values to memory locations via pointers, respectively. To support structured data types, the \textsf{Field} constraint enables field-sensitive analysis by facilitating precise differentiation between distinct fields of compound objects, thus enhancing analysis precision.

\subsection{Set-Constraint Based Pointer Analysis}
\label{sec:inclusion-based}

Set-constraint based pointer analysis is a widely-used formulation of inclusion-based pointer analysis, exemplified by Andersen's algorithm \cite{andersen1994}.
In this approach, each pointer variable is associated with a points-to set that conservatively represents all possible memory locations it may reference throughout program execution.
The pointer operations in the program are systematically encoded as set constraints, as illustrated in Table \ref{tab:constraint_solving_rule}, which are then solved iteratively to derive a sound approximation of the program's pointer relationships.

\begin{table*}[h!]

\centering

\caption{Set constraint-based points-to analysis rules.}
\renewcommand{\arraystretch}{1.3}
\footnotesize
\begin{tabular}{|c|c|c|}
\hline
Instruction & Constraint Type & Set Constraints \\ \hline
$\mathtt{p=\&o}$ & 
$\mathtt{p\xleftarrow{\textsf{Addr}}o}$ & 
$\mathtt{pts(p) = pts(p) \cup \{o\}}$ \\ \hline

$\mathtt{p=q}$ & 
$\mathtt{p\xleftarrow{\textsf{Copy}}q}$ & 
$\mathtt{pts(p) = pts(q) \cup pts(p)}$ \\ \hline

$\mathtt{p=\&q\!\rightarrow\! fld}$ & 
$\mathtt{p\xleftarrow{\textsf{Gep, fld}}q}$ & 
for each $\mathtt{o \in pts(q)}$ : $\mathtt{pts(p) = pts(p) \cup \{o.fld\}}$ \\ \hline

$\mathtt{p=*q}$ & 
$\mathtt{p\xleftarrow{\textsf{Load}}q}$ & 
for each $\mathtt{o \in pts(q)}$ : add 
$\mathtt{p\xleftarrow{\textsf{Copy}}o}$ \\ \hline

$\mathtt{*p=q}$ & 
$\mathtt{p\xleftarrow{\textsf{Store}}q}$ & 
for each $\mathtt{o \in pts(p)}$ : add 
$\mathtt{o\xleftarrow{\textsf{Copy}}q}$ \\ \hline

\end{tabular}
\label{tab:constraint_solving_rule}
\end{table*}

The five constraint types presented in Table \ref{tab:constraint_solving_rule} formalize the core semantics of pointer operations, establishing the foundation for solving points-to analysis problems. The \textsf{Addr} constraint establishes the fundamental mapping between pointer variables and their referenced memory objects. The \textsf{Copy} constraint enables the propagation of points-to relationships through direct assignments between variables. For memory operations, the \textsf{Load} and \textsf{Store} constraints operate by dereferencing the pointer operand and establishing a \textsf{Copy} constraint between the referenced object and the corresponding value operand. Additionally, the \textsf{GEP} (GetElementPtr) constraint facilitates field-sensitive analysis by precisely distinguishing between different fields of structured objects. When these constraints are applied iteratively within a set-constraint framework, they collectively construct a comprehensive model of the program's pointer relationships, which is essential for conducting precise and effective pointer analysis.

\subsection{Flow-Insensitive vs. Flow-Sensitive Pointer Analysis}
\label{sec:flow-insensitive}

Flow-insensitive pointer analysis (FI-PTA) computes a single points-to solution that remains invariant across all program points. This approach offers computational efficiency by abstracting away control flow considerations, enabling significant optimizations such as efficient propagation techniques~\cite{pereira_wave_2009}. However, this efficiency comes at the cost of precision, as the analysis frequently produces over-approximated points-to sets that can lead to false positives in downstream analyses~\cite{hardekopf_ant_2007}.
The implementation of FI-PTA typically employs a \textit{set-constraint graph} representation, where program variables form the nodes of the graph, and the different constraint types enumerated in Table~\ref{tab:constraint_solving_rule} manifest as edges between these nodes. This graph-based formulation transforms pointer analysis into a constraint-solving problem that can be efficiently resolved through iterative techniques.

\begin{figure*}
    \centering
    \includegraphics[width=\linewidth]{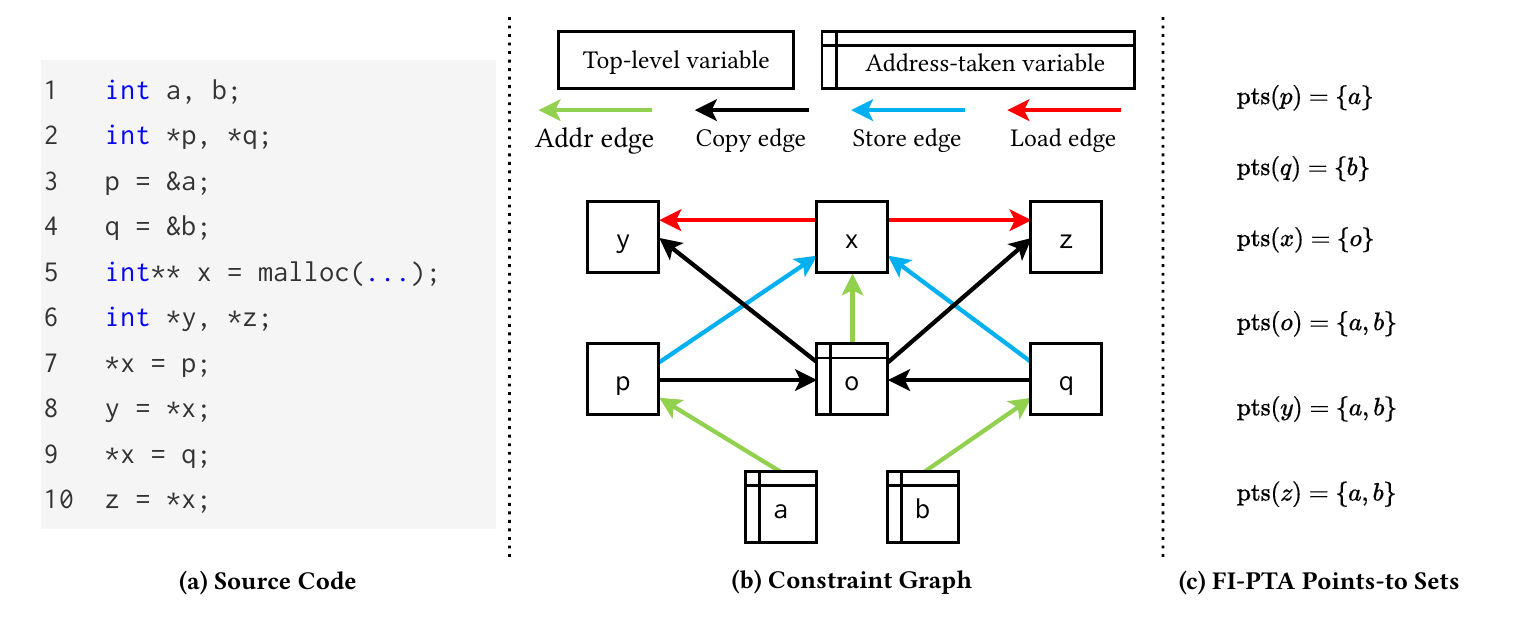}
    \vspace{-8mm}
    \caption{Flow-insensitive (constraint graph-based) pointer analysis.}
    \vspace{-5mm}
    \label{fig:background-combo1}
\end{figure*}

\begin{example}[{FI-PTA}]
Figure~\ref{fig:background-combo1}(a) presents a simple C program where top-level pointers $\mathtt{p}$, $\mathtt{q}$, and $\mathtt{x}$ point to $\mathtt{a}$, $\mathtt{b}$, and $\mathtt{o}$ respectively through \textsf{Addr} operations at program points $\lab_3$, $\lab_4$, and $\lab_5$. Subsequently, the program stores $\mathtt{p}$ to the memory location referenced by $\mathtt{x}$ at $\lab_7$ and loads from this location into $\mathtt{y}$ at $\lab_8$. This sequence is followed by storing $\mathtt{q}$ to the same memory location at $\lab_9$ and loading the value into $\mathtt{z}$ at $\lab_{10}$.

The flow-insensitive set-constraint graph corresponding to this code is illustrated in Figure~\ref{fig:background-combo1}(b). 
The analysis proceeds by iteratively applying the inference rules defined in Table~\ref{tab:constraint_solving_rule} to propagate points-to information throughout the graph without considering different program points. During constraint resolution, the black copy edges are dynamically added to reflect the propagation paths until a fixed point is reached—where no further changes to any points-to set occur. This approach, while computationally efficient, results in an over-approximation of the points-to sets for variables $\mathtt{o}$, $\mathtt{y}$, and $\mathtt{z}$, which are conservatively computed to contain both $\mathtt{a}$ and $\mathtt{b}$ without regard to the temporal ordering of assignments at different program points.
\end{example}

\begin{figure*}
    \centering
    \includegraphics[width=0.9\linewidth]{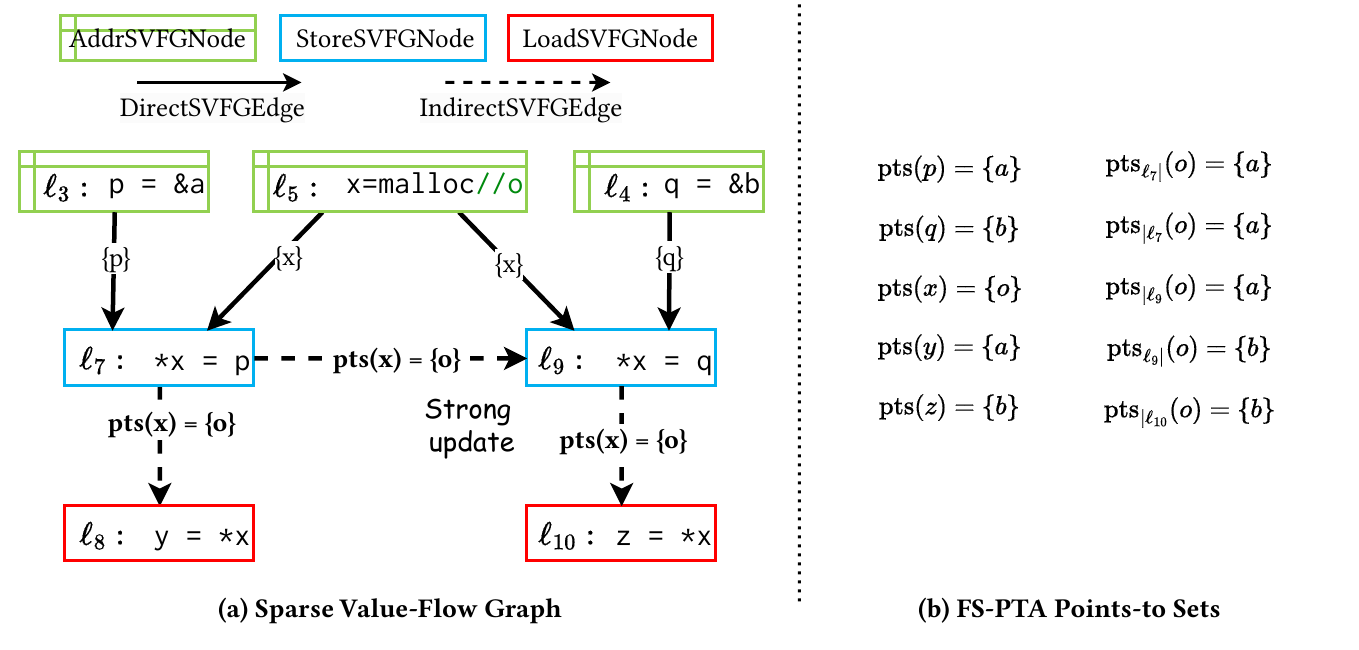}
    \vspace{-2mm}
    \caption{Flow-sensitive (sparse value-flow graph-based) pointer analysis. $\mathtt{pts_{|\lab_i}(p)}$ and $\mathtt{pts_{\lab_i|}(p)}$ represent the points-to set of $\mathtt{p}$ immediately before and after $\lab_i$, respectively.}
    \vspace{-5mm}
    \label{fig:background-combo2}
\end{figure*}

Flow-sensitive pointer analysis (FS-PTA) addresses the limitations of flow-insensitive approaches by considering the control flow and execution order of the program. Unlike FI-PTA, which computes a single points-to set for each variable across the entire program, FS-PTA maintains distinct points-to information at each program point, thereby respecting the sequence of instructions and enabling more precise modeling of pointer behaviors.

Hardekopf and Lin \cite{hardekopf_flow-sensitive_2011} introduce an efficient staged FS-PTA that leverages a preliminary conservative pointer analysis to construct a \emph{sparse value-flow graph} (SVFG). This graph, derived from the control-flow graph (CFG) of LLVM-IR, captures def-use relations between variable definitions through building interprocedural memory SSA~\cite{chow1996effective,hardekopf_flow-sensitive_2011, cc16}, serving as a foundation for subsequent, more precise flow-sensitive analysis by representing data dependencies that respect CFG execution order. During the flow-sensitive analysis phase, the approach maintains program-point-specific points-to sets for address-taken variables in non-SSA form, while top-level variables, already in SSA form, can be tracked globally without sacrificing precision. 

\begin{example}[{FS-PTA}]
Revisiting the code in Figure \ref{fig:background-combo1}, the SVFG in Figure \ref{fig:background-combo2} precisely captures value flows reflecting execution order. The value flow of $\mathtt{x}$ originates from $\lab_5:$\code{x = malloc}, establishing $\mathtt{x} \rightarrow \mathtt{o}$, and propagates to statements $\lab_7:$\code{*x = p} and $\lab_9:$\code{*x = q}. Similarly, $\mathtt{p}$ and $\mathtt{q}$ have value flows originating at $\lab_3:$\code{p = \&a} and $\lab_4:$\code{q = \&b}, respectively, which propagate to the corresponding store operations.

For the address-taken variable $\mathtt{o}$, three critical indirect value flows are captured: (1) from $\lab_7:$\code{*x = p} to $\lab_8:$\code{y = *x}, reflecting $\mathtt{y}$'s inheritance of $\mathtt{p}$'s value; (2) from $\lab_9:$\code{*x = q} to $\lab_{10}:$\code{z = *x}, indicating $\mathtt{z}$'s acquisition of $\mathtt{q}$'s value; and (3) from $\lab_7:$\code{*x = p} to $\lab_9:$\code{*x = q}, capturing the execution sequence between these statements. These def-use chains accurately model data flow at each program point.

By respecting execution order, FS-PTA precisely determines the different $\mathtt{o}$'s points-to set during program execution. We use the notation $\mathtt{pts}_{\mid\lab_i}$ and $\mathtt{pts}_{\lab_i\mid}$ to denote the points-to sets flowing into and out of label $\lab_i$, respectively. The analysis computes $\mathtt{pts}_{\lab_7\mid}(o) = \{a\}$, $\mathtt{pts}_{\mid\lab_7}(o) = \{a\}$ and $\mathtt{pts}_{\mid\lab_9}(o) = \{a\}$. After the strong update at $\lab_9$, we have $\mathtt{pts}_{\lab_9\mid}(o) = \{b\}$ and $\mathtt{pts}_{\mid\lab_{10}}(o) = \{b\}$. This precision is beyond the capability of FI-PTA, which conservatively merges all points-to information into a single points-to set $\{a,b\}$ for $\mathtt{o}$ across all program points. Note that the value $\{\mathtt{a}\}$ does not persist to $\lab_{10}$ due to the strong update at $\lab_9$ that overwrites the previous value.

\end{example}

SVFG-based FS-PTA achieves superior precision compared to set-constraint graph-based FI-PTA by maintaining distinct points-to sets at each program point. However, this precision improvement incurs computational overhead. Moreover, FS-PTA cannot fully exploit the algorithmic optimizations that have made FI-PTA computationally efficient, which are specifically tailored for set-constraint graph representations. This limitation creates a tension between precision and performance, motivating the development of an approach that effectively combines the precision advantages of FS-PTA with the computational efficiency of FI-PTA. Such an approach would enable scalable yet precise pointer analysis.

\subsection{Problem Formulation}
\label{sec:problem-formulation}

The core problem addressed in this work is to develop a novel approach to flow-sensitive pointer analysis that leverages the computational efficiency of set-constraint graph-based flow-insensitive analysis. Specifically, we aim to:

\begin{enumerate}[noitemsep, topsep=1pt, partopsep=1pt, listparindent=\parindent, leftmargin=*]
    \item Design a new set-constraint graph representation, called \emph{FSConsG}, that encodes flow-sensitivity within a set-constraint graph framework, preserving program point-specific pointer information without requiring an explicit control flow graph structure.
    
    \item Formulate a constraint resolution algorithm that maintains the precision benefits of flow-sensitive analysis — particularly the ability to distinguish points-to sets of address-taken variables at different program points — while utilizing efficient propagation techniques traditionally associated with flow-insensitive analysis.
    
\end{enumerate}

\section{Motivating Example}
\label{sec:motivation}

\begin{figure*}[t]
    \centering
    \includegraphics[width=0.9\linewidth]{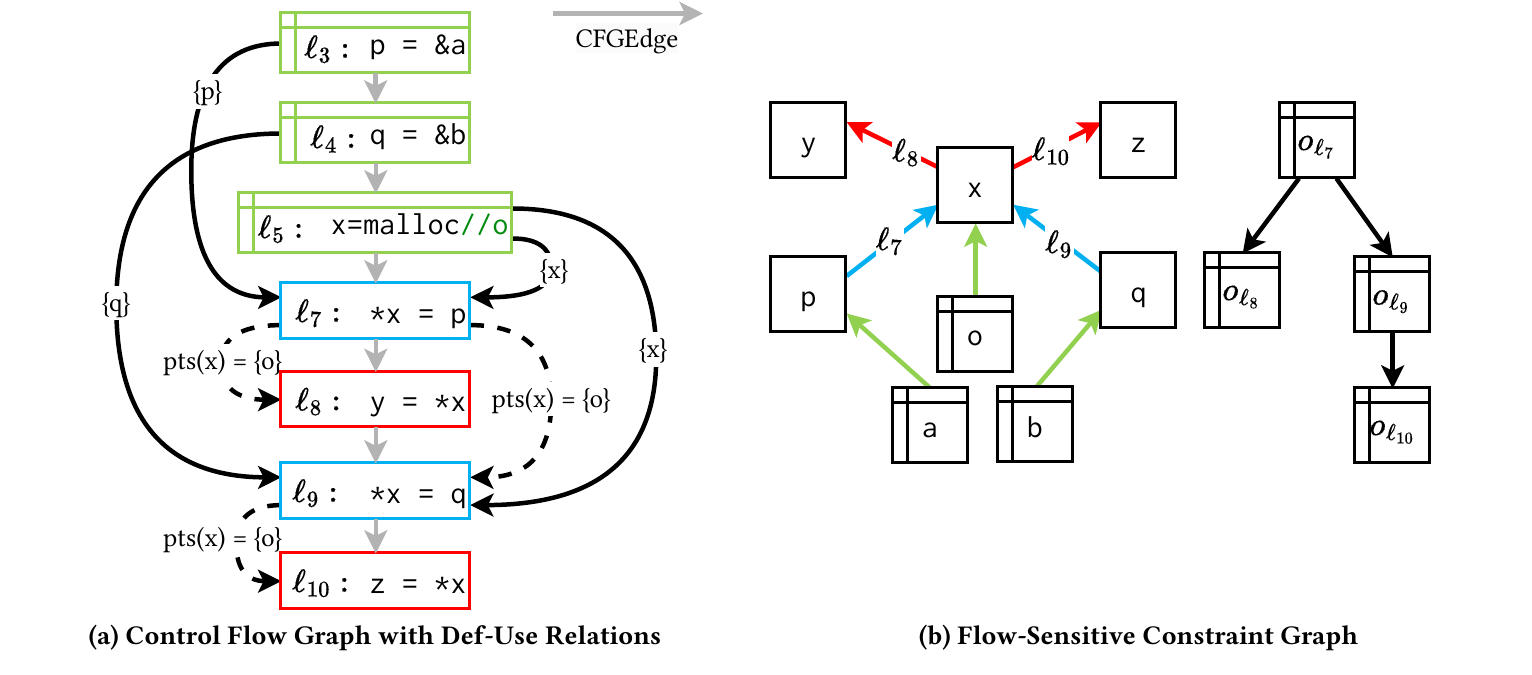}
    \vspace{-5mm}
    \caption{Construction of flow-sensitive constraint graph (FSConsG).}
    \label{fig:builder}
    \vspace{-5mm}
\end{figure*}

This section illustrates our \tool by revisiting the example in Figure~\ref{fig:background-combo1}, together with elucidating our approach and discussing its benefits compared to traditional SVFG-based FS-PTA techniques~\cite{hardekopf_flow-sensitive_2011,sui_value-flow-based_2018}.

\runinhead*{Construction of FSConsG.} Figure~\ref{fig:builder} illustrates how \tool constructs the flow-sensitive constraint graph (FSConsG), depicted in Figure~\ref{fig:builder}(b), based on the program's control flow graph (CFG) and pre-computed direct and indirect def-use relations, as shown in Figure~\ref{fig:builder}(a). 
The direct def-use relations between top-level variables are already explicitly defined in the program,
thus it is unnecessary to encode this information in the FSConsG.
The indirect def-use relations are derived from memory SSA forms constructed upon a preliminary pointer analysis. 
For each indirect relation, we augment the corresponding constraint edges (loads and stores) in the FSConsG with control flow information, generate versioned address-taken variables that correspond to dereferenced objects at these loads and stores, and establish copy edges between these versioned variables according to the pre-computed indirect def-use chains.

For instance, when processing the indirect def-use chain $\lab_7 \xrightarrow{\texttt{\{o\}}} \lab_8$, we consider the \textsf{Store} statement $\lab_7:$\code{*x=p} and the \textsf{Load} statement $\lab_8:$\code{y=*x}. We annotate the constraint edge $\store{p}{x}{\lab_7}$ with the control flow information from $\lab_7$ and the constraint edge $\load{x}{y}{\lab_8}$ with $\lab_8$. Based on the pre-computed points-to information $\texttt{pts(x)=\{o\}}$, we introduce a versioned address-taken variable $\mathtt{o_{\lab_7}}$ to represent the memory location dereferenced at $\lab_7$ and another versioned object $\mathtt{o_{\lab_8}}$ to represent $\lab_8$. Subsequently, to capture the indirect def-use relation $\lab_7 \xrightarrow{\texttt{\{o\}}} \lab_8$, we incorporate a \textsf{Copy} constraint $\mathtt{o_{\lab_7}} \xrightarrow{\textsf{Copy}} \mathtt{o_{\lab_8}}$ into the FSConsG.  

Analogously, for the indirect value-flow from $\lab_9$ to $\lab_{10}$ (i.e., $\lab_9 \xrightarrow{\texttt{\{o\}}} \lab_{10}$), we establish constraint edges $\store{q}{x}{\lab_9}$ and $\load{x}{z}{\lab_{10}}$, and add a \textsf{Copy} constraint $\mathtt{o_{\lab_9}} \xrightarrow{\textsf{Copy}} \mathtt{o_{\lab_{10}}}$ to the FSConsG. Furthermore, to capture the indirect def-use chain $\lab_7 \xrightarrow{\texttt{\{o\}}} \lab_9$, we incorporate the copy constraint $\mathtt{o_{\lab_7}} \xrightarrow{\textsf{Copy}} \mathtt{o_{\lab_9}}$ in the FSConsG.

By incorporating these versioned address-taken variables and establishing copy constraints between them according to indirect def-use chains, \tool effectively encapsulates the flow-sensitive semantics of address-taken variables at critical program points. This approach enables flow-sensitive pointer analysis to be performed using traditional constraint-graph-based techniques while preserving the precision benefits that flow-sensitivity offers. The resulting FSConsG thus serves as a unified representation that elegantly captures both points-to relations and their flow-sensitive properties in a single constraint framework.

\begin{figure*}[t]
    \centering
    \includegraphics[width=1\textwidth]{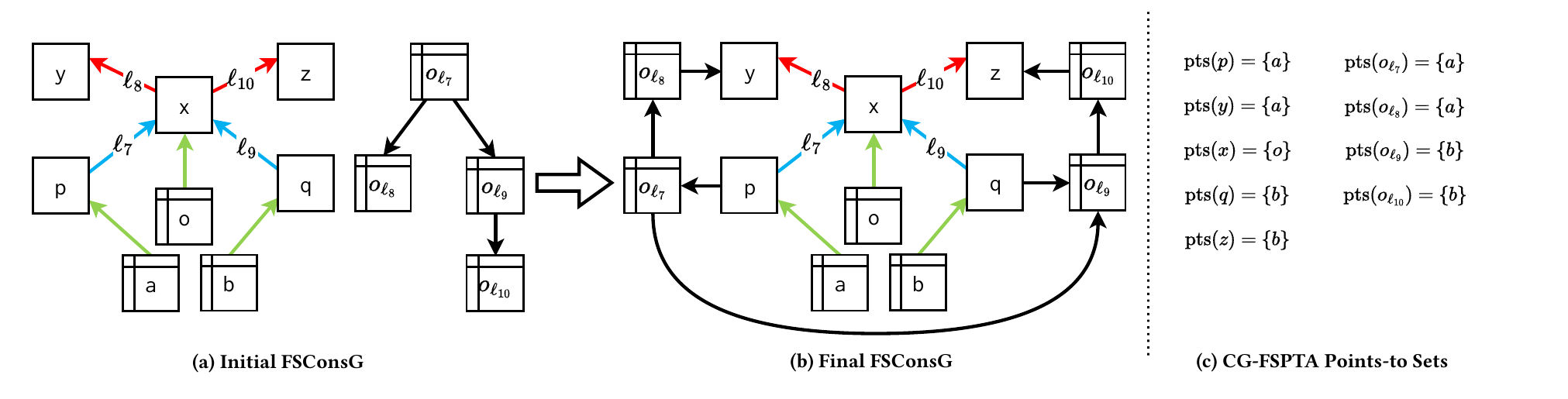}
    \vspace{-5mm}
    \caption{Flow-sensitive pointer analysis on FSConsG.}
    \vspace{-5mm}
    \label{fig:solver}
\end{figure*}

\runinhead*{FS-PTA on FSConsG.}
Figure~\ref{fig:solver} illustrates the process of flow-sensitive pointer analysis on FSConsG, which employs constraint solving according to the rules defined in Table~\ref{tab:constraint_solving_rule}. The figure depicts the updated FSConsG with newly added \textsf{Copy} edges and the points-to sets associated with each node. 
The analysis demonstrates that the resolved points-to sets of $\mathtt{y}$ and $\mathtt{z}$ yield the more precise flow-sensitive values $\mathtt{\{a\}}$ and $\mathtt{\{b\}}$ respectively, in contrast to the flow-insensitive result $\mathtt{\{a,b\}}$ obtained via FI-PTA. 
This enhanced precision stems from \tool's ability to differentiate the indirect value flows from $\mathtt{o_{\lab_7}}$ to $\mathtt{o_{\lab_8}}$ and from $\mathtt{o_{\lab_9}}$ to $\mathtt{o_{\lab_{10}}}$, thereby resolving the points-to sets of $\mathtt{y}$ and $\mathtt{z}$ with greater precision. Note that the points-to set of $\mathtt{o_{\lab_7}}$ does not propagate to $\mathtt{o_{\lab_9}}$ due to the strong update semantics~\cite{strongupdate} applied to the \textsf{Copy} constraint.

\runinhead*{Benefits.}
The FSConsG representation offers efficient structural simplification compared to a traditional SVFG. For instance, while the variable $x$ appears in multiple program statements ($\lab_5, \lab_7,\lab_8,\lab_9,\lab_{10}$) in the SVFG, it corresponds to exactly one node in the FSConsG.
More importantly, FSConsG represents constraints solely through edges, whereas SVFG requires maintaining constraints on both nodes and edges. This design choice reduces the redundant number of constraints to be maintained during analysis, enabling our approach to achieve the precision of flow-sensitive pointer analysis with the computational efficiency of constraint graph-based techniques~\cite{pereira_wave_2009,chang_fast_2019}.

\section{Flow-Sensitive Constraint Graph (FSConsG)}

This section presents the definition and construction method of our Flow-Sensitive Constraint Graph (FSConsG).

\subsection{Formal Definition of FSConsG}

A Flow-Sensitive Constraint Graph, $\text{FSConsG}$, is formally defined as a directed graph $G_{FS} = (\mathcal{V}_{FS}, E_\textit{FS})$. We next detail the nodes and edges of FSConsG.

\paragraph{Set of Nodes ($\mathcal{V}_{FS}$)}

The node set $\mathcal{V}_{FS}$ consists of top-level variables and program-point-versioned address-taken variables:
$\mathcal{V}_{FS} = \mathcal{P} \cup \mathcal{O}_{\mathcal{L}}$.

Here, $\mathcal{P}$ denotes the set of top-level pointer variables represented as SSA values.
Let $\mathcal{O}$ be the set of abstract memory objects and $\mathcal{L}$ the set of program
points (CFG node identifiers).
We define a partial versioning function
\[
\nu : \mathcal{O} \times \mathcal{L} \rightharpoonup \mathcal{O}_{\mathcal{L}},
\]
with a domain characterized as
\[
\nu(o,\ell)\ \text{is defined}
\;\Longleftrightarrow\;
\ell \in \mathrm{IndDef}(o) \cup \mathrm{IndUse}(o),
\]
where $\mathrm{IndDef}(o)$ and $\mathrm{IndUse}(o)$ denote the sets of program points at which
object $o$ is indirectly defined and indirectly used, respectively.
When defined, $\nu(o,\ell) = o_{\ell}$ denotes the flow-sensitive version of object $o$
at program point $\ell$.

The set of flow-sensitive object nodes is
\[
\mathcal{O}_{\mathcal{L}} \;\triangleq\; \{\, o_{\ell} \mid \nu(o,\ell)\ \text{is defined} \,\}.
\]
Specifically, for each indirect def--use edge
\[
\ell_i \xrightarrow{\mathrm{pts}(o)} \ell_j,
\]
where $\ell_i \in \mathrm{IndDef}(o)$ and $\ell_j \in \mathrm{IndUse}(o)$, we create two
FSConsG nodes $o_{\ell_i}$ and $o_{\ell_j}$.
Here, $o_{\ell_i}$ represents the definition-site version of $o$, whereas
$o_{\ell_j}$ represents the corresponding use-site version.


\paragraph{Set of Edges (${E}_\textit{FS}$)}

Similar to flow-insensitive constraint graphs, an FSConsG also contains the five basic types of constraint edges, i.e., \texttt{Addr}, \texttt{Copy}, \texttt{Store}, \texttt{Load}, and \texttt{Gep}.
The difference is that each edge is annotated with an index $\ell$, denoting the execution points of the represented statement in the program.
In particular, there are exemptions for \texttt{Addr}, \texttt{Copy} and \texttt{Gep} edges, which are detailed in Section \ref{sec:top-level}.

\paragraph{\textbf{Remark} (Handling arbitrary pointer levels). Although the above definition focuses on single-level dereferences in \texttt{Load} and \texttt{Store} edges, FSConsG naturally supports double and higher-level pointers. As described in Section~\ref{sec:LLVMIR}, LLVM-IR decomposes multi-level pointer operations into sequences of single-level \texttt{load}/\texttt{store} instructions, each involving a top-level variable (represented as a single SSA node in $\mathcal{P}$) and an address-taken variable (versioned in $\mathcal{O}_\mathcal{L}$ at the corresponding program points). Each such instruction maps to exactly one FSConsG edge. Consequently, the FSConsG framework uniformly handles arbitrary levels of pointer indirection without any special treatment.}

\subsection{Construction of FSConsG}
\label{sec:builder}

Our flow-sensitive constraint graph is constructed on top of the result of the conservative flow-insensitive pointer analysis \cite{chang_fast_2019} and a control flow graph (CFG) \cite{hardekopf_flow-sensitive_2011}, constructed in preprocessing.
In particular, in the CFG, we follow the technique of \cite{sui_sparse_2016} to compute the def-use relations of variables.

\subsubsection{Converting Program Statements}
\label{sec:top-level}
The first step of constructing an FSConsG is to convert the statements in the CFG nodes into FSConsG edges.
This establishes the basic structure of our FSConsG, a digraph describing the operations of top-level variables via edges.
Since the flow-insensitive analysis and the construction of CFG in the preprocessing stage are both conducted based on the LLVM-IR, top-level variables are already in the SSA form.
Moreover, the statements of CFG nodes involve only top-level variables, except for the \texttt{Addr} statements, which allocate an abstract memory object to a top-level variable.
The rules of converting the five statements, involving top-level operations, into edges are displayed in Figure \ref{fig:statements}.
The following part details each conversion rule.

\paragraph{Address}
The \texttt{Addr} CFG nodes capture LLVM-IR's \texttt{alloca} instructions from abstract memory objects to top-level variables.
We use the rule \texttt{BuildAddr} in Figure \ref{fig:statements} to facilitate the transformation from \texttt{Addr} CFG nodes to \texttt{Addr} edges in FSConsG. Specifically, a \texttt{Addr} statement $\lab_i:$ \texttt{p = \&o}, which captures an LLVM-IR's \texttt{alloca} instruction, is converted into two nodes $p$ and $o$, and an edge $\addr{o}{p}{}$ in the FSConsG.

\paragraph{Copy}
The \texttt{Copy} CFG nodes capture LLVM-IR's \texttt{copy} instructions.
We use the rule \texttt{BuildCopy} in Figure \ref{fig:statements} to facilitate the transformation from \texttt{Copy} CFG nodes to \texttt{Copy} edges in FSConsG.
Specifically, a CFG node containing a \texttt{Copy} statement $\lab_i:$ \texttt{q = p}, which captures an LLVM-IR's \texttt{copy} instruction, is converted into two nodes $p$ and $q$, and an edge $\cpy{p}{q}{}$ in the FSConsG.

\paragraph{Gep}
For field-sensitive analysis, the \texttt{Gep} CFG nodes capture LLVM-IR's \texttt{getelementptr} instructions about field access operations.
We use the rule \texttt{BuildGep} in Figure \ref{fig:statements} to facilitate the transformation from \texttt{Gep} CFG nodes to \texttt{Gep} edges in FSConsG.
Specifically, a \texttt{Gep} statement $\lab_i:$ \texttt{q = \&p.fld}, which captures an LLVM-IR's \texttt{getelementptr} instruction, is converted into two nodes $p$ and $q$, and an edge $\gep{p}{q}{fld}$ in the FSConsG.

\paragraph{Store} 
LLVM-IR uses indirect instructions \texttt{Store} and \texttt{Load} to module the accesses and operations of address-taken variables.
The \texttt{Store} nodes in a CFG capture LLVM-IR's \texttt{store} instructions. 
We use the rule \texttt{BuildStore} in Figure \ref{fig:statements} to facilitate the transformation from \texttt{Store} CFG nodes to \texttt{Store} edges in FSConsG.
Specifically, a \texttt{Store} statement $\lab_i:$ \texttt{*q = p}, which captures an LLVM-IR's \texttt{store} instruction, is converted into two nodes $p$ and $q$, and an edge $\store{p}{q}{\lab_i}$ in the FSConsG.

\paragraph{Load} 
The \texttt{Load} nodes in a CFG capture LLVM-IR's \texttt{load} instructions. 
We use the rule \texttt{BuildLoad} in Figure \ref{fig:statements} to facilitate the transformation from \texttt{Load} CFG nodes to \texttt{Load} edges in FSConsG.
Specifically, a \texttt{Load} statement $\lab_i:$ \texttt{q = *p}, which captures an LLVM-IR's \texttt{load} instruction, is converted into two nodes $p$ and $q$, and an edge $\load{p}{q}{\lab_i}$ in the FSConsG.

\begin{figure*}[h!]
  \centering
  \hspace{-3mm}
  \begin{tabular}{l}
    \ruledef{\rulename{BuildAddr}}{\texttt{CFGNode } \lab_i : \texttt{p = \&o}}{ \addr{o}{p}{\lab_i}\in E_\textit{FS}}
  \end{tabular}
  \vspace{2mm}
  
  \hspace{-3mm}
  \begin{tabular}{l@{\hspace{5mm}}l}
    \ruledef{\rulename{BuildCopy}}{\texttt{CFGNode } \lab_i : \texttt{q = p}}{\cpy{p}{q}{\lab_i}\in E_\textit{FS}}
    &
    \ruledef{\rulename{BuildGep}}{\texttt{CFGNode } \lab_i : \texttt{q = \&p.fld}}{\gep{p}{q}{fld}\in E_\textit{FS}}
  \end{tabular}
  \vspace{2mm}

  \hspace{-3mm}
  \begin{tabular}{l@{\hspace{5mm}}l}
    \ruledef{\rulename{BuildStore}}{\texttt{CFGNode } \lab_i : \texttt{*q = p}}{\store{p}{q}{\lab_i}\in E_\textit{FS}}
    &
    \ruledef{\rulename{BuildLoad}}{\texttt{CFGNode } \lab_i : \texttt{q = *p}}{\load{p}{q}{\lab_i}\in E_\textit{FS}}
  \end{tabular}
  
  \caption{Rules for converting top-level operations. Here, $e\in E_\textit{FS}$ denotes adding an edge $e$ to the FSConsG.}
  \vspace{-5mm}
  \label{fig:statements}
\end{figure*}

\paragraph{Maintaining control flow information}
Note that in the rules of Figure \ref{fig:statements}, each FSConsG edge is also annotated with $\lab_i$, the index of the converted CFG node (e.g., $\store{p}{q}{\lab_i}$).
We use this annotation to capture the execution order manifested in CFG nodes and preserve the flow sensitivity of top-level operations.
This is imperative in flow-sensitive constraint solving on FSConsG, which is detailed in Section \ref{sec:solver}.
Notably, the top-level variables are already in the SSA form in LLVM-IR, which means that the statements involving only the points-to sets of top-level variables (i.e., \texttt{Addr} and \texttt{Copy}) in the CFG are inherently flow-sensitive.
Thus, we do not need to maintain the corresponding CFG node indices for \texttt{Addr} and \texttt{Copy} edges in an FSConsG.

\subsubsection{Maintaining Def-Use Relations}
\label{sec:address-taken}

The methods of Section \ref{sec:top-level} convert a CFG into a set-constraint graph consisting of discrete edges.
To make a complete FSConsG, we further correlate the discrete edges via def-use relations, producing a connected graph while maintaining flow-sensitivity.




In a CFG, a def-use edge between two nodes denotes that some common variable(s) are defined in a program point while redefined/used in another program point.
Def-use edges include direct ones (the def-use relations of top-level variables) and indirect ones (the def-use relations of address-taken variables), and can be calculated by integrating the result of flow-insensitive preanalysis into the CFG \cite{sui_sparse_2016}.

For example, Figure \ref{fig:builder}(a) is a CFG with def-use, the gray edges denote control flows, and the solid and dashed black edges denote direct and indirect def-use edges, respectively.
The direct def-use edge from node $\lab_3$ to node $\lab_7$, annotated with $\{p\}$, indicates that $p$ is defined at $\lab_3$ (via \texttt{p = \&a}) and used at $\lab_7$ (via \texttt{*x = p}).
Relatively, the indirect def-use edge from $\lab_7$ to $\lab_8$, annotated with $\textit{pts}(x) = \{o\}$, indicates that $o$ is (re)defined at $\lab_7$ (via \texttt{*x = p}) and used in $\lab_8$ (via \texttt{y = *x}).
In particular, $\textit{pts}(x) = \{o\}$ is calculated in the flow-insensitive preanalysis.

Different from an SVFG that is constructed via connecting CFG nodes with def-use edges and removing the original CFG edges, our FSConsG directly embeds the def-use relations into a set-constraint graph.
The following details how we handle top-level and address-taken variables, respectively.


\paragraph{Avoiding top-level collisions}
\label{sec:avoid}
Collisions of FSConsG nodes will emerge when several CFG nodes involve the same top-level variable.
A brute-force approach is to separate the common variable into several versions, corresponding to the CFG nodes containing the variable, then connect the versions with \texttt{Copy} edges based on the control flow.
For example, in Figure \ref{fig:builder}(a), consider the two CFG nodes $\lab_3: \texttt{p = \&a}$ and $\lab_7: \texttt{*x = p}$ that are connected by a direct def-use edge annotated with $\{p\}$. 
The brute-force approach separates $p$ into to versions, i.e., $p_1$ and $p_2$, constructing two edges $\addr{a}{p_1}{}$ and $~\store{p_2}{x}{\lab_7}$, then using a \texttt{Copy} edge to connect $p_1$ and $p_2$, forming a path $a \xrightarrow{\texttt{Addr}} p_1 \xrightarrow{\texttt{Copy}} p_2 \xrightarrow{\texttt{Store},~\lab_7} x$.

We can easily find the redundancy caused by the brute-force method, i.e., $p_1 \xrightarrow{\texttt{Copy}} p_2$.
In fact, in a CFG construct from LLVM-IR, all top-levels are already in the SSA form, hence, they do not need to be separated into versions.
Different from the brute-force method, our technique builds a \textit{symbol table} along with the construction process of FSConsG. 
Specifically, when converting a CFG node $\lab$, if a variable $v$ involved in $\lab$ is not in the symbol table, then record it into the symbol table; otherwise, we directly use $v$ to connect the newly transformed FSConsG edge. Thus, the top-level variables do not need to be separated into versions.
Consequently, in the above example, the connected $\ell_3$ and $\ell_7$ will be transformed into $a \xrightarrow{\texttt{Addr}} p \xrightarrow{\texttt{Store},~\lab_7} x$.



\paragraph{Address-taken versioning.}
Based on the flow-insensitive preanalysis, we create versioned address-taken FSConsG nodes according to the information on the indirect def-use edges.
Specifically, for an indirect def-use edge $\lab_i \xrightarrow{pts(o)} \lab_j$, we create two FSConsG nodes $o_{\lab_i}$ and $o_{\lab_j}$ of $o$, and connect a \texttt{Copy} edge from $o_i$ to $o_j$, indicating that the points-to information of $o_i$ defined in $\lab_i$ will directly flow to $o_j$ defined in $\lab_j$.
By transforming the indirect def-use edges one by one, we create a series of \textit{versioned} address-taken nodes that are connected with \texttt{Copy} edges in FSConsG.
In this way, preserving the flow sensitivity of address-taken variables via chains of address-taken nodes, and, hence, converting a flow-insensitive constraint graph into a flow-sensitive one.

\begin{example}[\textit{Address-Taken Versioning}]
    Consider the indirect def-use edge $\lab_7 \xrightarrow{\{o\}} \lab_8$ in Figure \ref{fig:builder}(a). Two address-taken FSConsG nodes $o_{\lab_7}$ and $o_{\lab_8}$, and a \texttt{Copy} edge $o_{\lab_7} \xrightarrow{\texttt{Copy}}~o_{\lab_8}$ are created and added to the FSConsG.
    Similarly, $o_{\lab_7} \xrightarrow{\texttt{Copy}} o_{\lab_9}$ and $o_{\lab_9} \xrightarrow{\texttt{Copy}} o_{\lab_{10}}$ are created and added to the FSConsG, according to $\lab_7 \xrightarrow{\{o\}} \lab_9$ and $\lab_9 \xrightarrow{\{o\}} \lab_{10}$, respectively, as seen in Figure \ref{fig:builder}(b).
\end{example}





\section{FSConsG-Based Solver}
\label{sec:solver}

This section introduces an FSConsG-based constraint solver for solving flow-sensitive pointer analysis.
The solver is presented in Algorithm \ref{algo:solver}, which accepts an LLVM-IR and returns the flow-sensitive pointer analysis result, in the form of points-to sets.

\subsection{Solver Framework}

The solving procedure is comprised of three stages - preprocessing (lines \ref{line:preanalysis}), FSConsG construction (lines \ref{line:fsconsg}, and points-to-set solving (lines \ref{line:addr}--\ref{line:repeat-end}).
The following goes into the details.

The preprocessing stage focuses on three issues: 
(1) constructing a CFG and a flow-insensitive constraint graph according to the input LLVM-IR;
(2) running flow-insensitive pointer analysis on the constraint graph;
(3) determining def-use edges on the CFG based on the result of flow-insensitive preanalysis.
In particular, we use the Wave solver \cite{pereira_wave_2009} to perform the flow-insensitive pointer analysis and Sui et al.'s approach \cite{sui_sparse_2016} to determine def-use edges.

The process of constructing FSConsG follows the instructions demonstrated in Section \ref{sec:builder}.

We adopt the staged solving strategy of Pereria et al. \cite{pereira_wave_2009,lei2019fast} to solve FSConsG.
Specifically, the FSConsG solver is initialized with points-to sets calculated from \texttt{Addr} edges (line \ref{line:addr}), and an empty worklist (line \ref{line:w}).
The solving process (lines \ref{line:rep2}--\ref{line:rep1}) is conducted under iterations, with points-to-set propagations (lines \ref{line:stg1-start}--\ref{line:stg1-end}) and edge insertions (lines \ref{line:stg2-start}--\ref{line:stg2-end}) categorized into two different stages.
The state-of-the-art graph simplification techniques, cycle elimination \cite{nuutila_finding_1994,tarjan1972depth} and graph folding \cite{cook1979graph,lei2023recursive}, are applied to the graph at the beginning of each iteration, as seen in line \ref{line:gsimp}.

By utilizing Nuutila's approach \cite{nuutila_finding_1994}, 
the \textsf{SCC()} method can return a topological order of the nodes in the graph, which is used for the following constraint-solving.
At the end of each iteration, to handle indirect call relations, our solver updates the call graph (line \ref{line:callgraph}) and decides whether to start a new iteration according to whether the set of edges $E$ is changed (line \ref{line:reanalyze}).

\begin{algorithm}
\caption{\tool Solver}
\label{algo:solver}

\begin{algorithmic}[1]
\REQUIRE LLVM-IR
\ENSURE Result of flow-sensitive analysis as points-to sets
\STATE \label{line:preanalysis} Preanalysis: construct CFG, flow-insensitive analysis, and def-use edges
\hfill 
\STATE \label{line:fsconsg} Construct \textit{FSConsG}$\langle V,E \rangle$ based on the CFG with explicit def-use edges
\hfill 

\STATE \label{line:addr} $\forall \addr{o}{p}{\ell_i} \in E_{Addr}$: $\pts(p) \leftarrow \{o\}$
\STATE \label{line:w} \textbf{let} $W$ be a worklist, initially empty
\REPEAT \label{line:rep1}
    \STATE \label{line:rep2} $E_\textit{old} \leftarrow E$
    \STATE \label{line:gsimp} \textit{topologicalOrder} $\leftarrow$ \textsf{SCC()}
    \hfill 
    \FOR{\label{line:stg1-start} each $v \in \textit{topologicalOrder}$}
        \STATE \label{line:copy} $\forall \cp{v}{w}$: $\pts(w) \cupeq \pts(v)$; 
        \IF{$\pts(w)$ changed}
        \STATE $W \leftarrow W \cup \{w\}$
        \ENDIF
        \STATE \label{line:gep} $\forall \gep{v}{w}{fld}$: $\pts(w) \cupeq \{o_{fld} \mid o \in \pts(v)\}$; 
        \IF{$\pts(w)$ changed} 
        \STATE $W \leftarrow W \cup \{w\}$
        \ENDIF
    \ENDFOR \label{line:stg1-end} 
    \WHILE{\label{line:stg2-start} $W \neq \emptyset$}
        \STATE select and remove a node $v$ from $W$
        \FOR{each $o \in \pts(v)$}
            \STATE $\forall \store{u}{v}{\ell_i} \in E$: 
                \IF{\textsf{IsStrongUpdate()}}
                    \STATE \label{line:su} $\pts(o_{\ell_i}) \leftarrow \emptyset$; \label{line:store1} $E \cupeq \{\cp{u}{o_{\ell_i}}\}$; $W \leftarrow W \cup \{u\}$
                \ELSIF{$\cp{u}{o_{\ell_i}}\notin E$}
                    \STATE \label{line:store2} $E \cupeq \{\cp{u}{o_{\ell_i}}\}$; $W \leftarrow W \cup \{u\}$
                \ENDIF
            \STATE $\forall \load{v}{w}{\ell_i}\in E$: 
                \IF{$\cp{o_{\ell_i}}{w} \notin E$}
                    \STATE \label{line:load} $E \cupeq \{\cp{o_{\ell_i}}{w}\}$; $W \leftarrow W \cup \{o_{\ell_i}\}$
                \ENDIF
        \ENDFOR
    \ENDWHILE \label{line:stg2-end}
    \STATE \label{line:callgraph} \textsf{UpdateCallgraph()} \hfill 
    \IF{$E \neq E_\textit{old}$}
        \STATE \label{line:reanalyze} \textit{reanalyze} = \textbf{true}
    \ELSE
        \STATE \textit{reanalyze} = \textbf{false}
    \ENDIF
\UNTIL{\textit{reanalyze} = \textbf{false} \label{line:repeat-end}}

\end{algorithmic}
\end{algorithm}

\subsection{Constraint Solving}

This part details how our technique solves flow-sensitive set constraints on the FSConsG.
\texttt{Addr}, \texttt{Copy} and \texttt{Gep} edges are solved by the same schemes of flow-insensitive analysis.
Namely, each \texttt{Addr} edge adds its source to its target (line \ref{line:addr}), each \texttt{Copy} edge adds its points-to objects to its target's points-to set (line \ref{line:copy}), and each \texttt{Gep} edge adds the fields of its points-to objects to its target's points-to set (line \ref{line:gep}).
\texttt{Store} and \texttt{Load} edges involve indirect def-use relations, hence they are solved differently to maintain the flow-sensitivity of address-taken variables.

\paragraph{Processing \texttt{Store} edges}
Different from flow-insensitive analysis, the \texttt{Store} edges in an FSConsG contain the information of CFG node index, denoting the position of the \texttt{Store} instruction in the CFG. 
As illustrated in Section \ref{sec:address-taken}, we use CFG node indices to relate versions of address-taken variables (nodes) and the instructions defining (\texttt{Store} edges) or using them (\texttt{Load} edges).
Therefore, when processing an \texttt{Store} edge, e.g., $\store{u}{v}{\ell_i}$ where the points-to object of $v$ is (re)defined, we traverse $\pts(v)$.
And, for each $o \in \pts(v)$,
we use the CFG node index, i.e., $\ell_i$, to locate the corresponding version of $o$, i.e., $o_{\ell_i}$, and connect the corresponding $\texttt{Copy}$ edges to $o_{\ell_i}$, as seen in lines \ref{line:store1} and \ref{line:store2}.
In particular, our technique also checks strong updates \cite{hardekopf_flow-sensitive_2011} when processing a \texttt{Store} edge. 
If the \texttt{Store} edge is a strong update, then we empty $\pts(o_{\ell_i})$, as seen in line \ref{line:su}.

\paragraph{Processing \texttt{Load} edges}
Similar to processing \texttt{Store} edges, 
when processing a \texttt{Load} edge, e.g., $\load{v}{w}{\ell_i}$ where the points-to object of $v$ is used, we traverse $\pts(v)$.
And, for each $o \in \pts(v)$,
we use the CFG node index, i.e., $\ell_i$, to locate the corresponding version of $o$, i.e., $o_{\ell_i}$, and connect the corresponding $\texttt{Copy}$ edges from $o_{\ell_i}$ to the target $w$, as seen in line \ref{line:load}.

\paragraph{Soundness and Precision Guarantee}
\tool guarantees its soundness and precision by producing points-to-set results identical to VSFS, which is embedded in the graph construction and constraint solving processes.
In the graph construction process, 
since both techniques are based on  LLVM-IR, where the top-level variables are already in the SSA form, the top-level variables in SVFGs and FSConsGs are naturally flow-sensitive.
For address-taken variables, SVFG uses def-use edges connecting SVFG nodes to facilitate different points-to sets of an address-taken variable in different SVFG nodes.
Correspondingly, \tool separates address-taken variables into different versions, represented as FSConsG nodes, and uses \texttt{Copy} edges to connect them, strictly according to the def-use edges (Section \ref{sec:address-taken}).
Thus, the different versions of an address-taken node in an FSConsG correspond to different points-to sets of an address-taken variable in an SVFG.
In the solving process, when processing an indirect SVFG node (\texttt{Store} and \texttt{Load}), VSFS uses indirect SVFG edges to distinguish different versions of the involved address-taken variables.
Correspondingly, when processing an indirect FSConsG edge, \tool uses the CFG node number on the edge to locate and process the correct version of the corresponding address-taken nodes.
Therefore, \tool and VSFS produce the same points-to-set results.

\section{Evaluation}
\label{Experiment}

This section evaluates \tool's performance in flow-sensitive pointer analysis through a comparative analysis with VSFS~\cite{barbar_object_2021}, a state-of-the-art flow-sensitive pointer analysis framework based on CFG. Our experimental evaluation demonstrates that \tool achieves significant performance improvements: \tool achieves a \textbf{1.93\stimes} average execution time speedup (up to \textbf{2.91\stimes}) and \textbf{23.99\%} memory reduction (up to \textbf{54.31\%}) when utilizing Andersen's \wave propagation algorithm \cite{pereira_wave_2009}, and a \textbf{7.27\stimes} average speedup (up to \textbf{20.07\stimes}) and \textbf{33.05\%} memory reduction (up to \textbf{61.07\%}) when implementing the \sfr technique \cite{chang_fast_2019}. Importantly, these substantial performance gains are achieved while maintaining precision equivalence with VSFS in terms of points-to set computation.

\subsection{Experimental Setup}

\runinhead*{Datasets.} Our evaluation employs the SPEC CPU 2017 Benchmark Suite~\cite{spec2017}, which comprises a diverse collection of real-world C/C++ applications spanning multiple domains and encompassing over 1 million lines of code. We compiled these programs into LLVM bitcode files (version 16) for analysis. Table~\ref{tab:benchmark} presents detailed statistics for each benchmark. The suite includes: \texttt{lbm} (fluid dynamics), \texttt{mcf} (combinatorial optimization for route planning), \texttt{namd} (molecular dynamics simulation), \texttt{deepsjeng} (artificial intelligence: alpha-beta tree search for chess), \texttt{leela} (artificial intelligence: Monte Carlo tree search for Go), \texttt{nab} (nucleic acid simulation), \texttt{xz} (general data compression), \texttt{x264} (H.264/AVC video encoding), \texttt{omnetpp} (discrete event simulation for computer networks), \texttt{povray} (ray tracing renderer), \texttt{cactus} (physics: numerical relativity), and \texttt{imagick} (image manipulation). These benchmarks were selected to represent a diverse range of program sizes and complexities, enabling a comprehensive evaluation of our approach across various scenarios.

\runinhead*{Implementation.}
We conducted our experiments on a Rocky Linux 8.10 server equipped with 28-core Intel Xeon E5-2690v4 2.6GHz CPUs and 252 GB of memory.
Our implementation builds upon LLVM version 16.0.0 \cite{lattner2004llvm}, constructing both the interprocedural control flow graph \cite{schiebel2024scaling} and ConsG based on LLVM's intermediate representation (LLVM-IR).
In LLVM-IR, program functions are represented as global variables, which we model as address-taken variables during analysis.
The analysis proceeds in several phases. First, we perform a flow-insensitive pointer analysis \cite{andersen1994} as a preprocessing step. This initial analysis enables the construction of interprocedural memory SSA form \cite{chow1996effective,hardekopf_flow-sensitive_2011}, which provides the foundation for subsequently building our FSConsG representation. 
For comparative evaluation, we utilize VSFS's~\cite{barbar_object_2021} original open-source implementation for SVFG-based flow-sensitive pointer analysis as our baseline~\cite{cc16}. We then evaluate our \tool implementation, which integrates the FSConsG representation with two established inclusion-based pointer analysis techniques: Andersen's \wave propagation algorithm \cite{pereira_wave_2009} and the more recent \sfr technique \cite{chang_fast_2019}.

\begin{table*}[t]
  \centering
  \small
  \caption{Statistics of benchmarks from SPEC CPU 2017~\cite{spec2017}. KLOC represents thousands of lines of code. \#Pointers, \#Gep, \#Load, and \#Store denote the number of pointer variables, getelementptr instructions (field accesses), load instructions, and store instructions in the LLVM IR representation, respectively.}
  \label{tab:benchmark}
    \begin{tabular}{ ll >{\raggedleft\arraybackslash}p{0.15\textwidth} rrrr }
        \toprule
        \textbf{No.} & \textbf{Project} & \textbf{KLOC} & \textbf{\#Pointers} & \textbf{\#Gep} & \textbf{\#Load} & \textbf{\#Store} \\
          \midrule
          1&\texttt{lbm}   & 1     & 4,131  & 437   & 256   & 313  \\
          2&\texttt{mcf}   & 3     & 7,979  & 1,201  & 858   & 709  \\
          3&\texttt{namd}  & 8     & 332,018  & 42,700  & 33,159  & 15,949  \\
          4&\texttt{deepsjeng} & 10    & 28,517  & 3,392  & 2,983  & 2,188  \\
          5&\texttt{leela} & 21    & 68,078  & 8,098  & 5,263  & 4,063  \\
          6&\texttt{nab}   & 24    & 55,925  & 6,203  & 6,617  & 4,657  \\
          7&\texttt{xz}    & 33    & 46,880  & 5,348  & 3,941  & 3,959  \\
          8&\texttt{x264}  & 96    & 258,037  & 67,934  & 41,617  & 34,728  \\
          9&\texttt{omnetpp} & 134   & 494,762  & 51,385  & 47,757  & 27,090  \\
          10&\texttt{povray} & 170   & 279,494  & 35,867  & 31,161  & 25,729  \\
          11&\texttt{cactus} & 257   & 209,738  & 39,117  & 32,039  & 14,731  \\
          12&\texttt{imagick} & 259   & 507,367  & 55,404  & 51,145  & 32,558  \\
          \midrule
          \textbf{\textit{Total}} & & 1,016  & 2,288,926 & 312,086 & 260,796 & 166,673 \\
    \bottomrule
    \end{tabular}%
    \vspace{-3mm}
\end{table*}%

\runinhead*{Evaluation Metrics.} Our comparative assessment focuses on two fundamental metrics: the number of constraints within the respective graph representations and the total execution time. 
The constraints of a graph constitute the structural foundation upon which flow-sensitive pointer analysis depends.
Specifically, SVFG constraints encompass both node-level elements (instructions) and edge-level connections (def-use relations), whereas FSConsG exclusively contains edge-level constraints (as specified in Table~\ref{tab:constraint_solving_rule}).
The constraint count metric directly quantifies the structural efficiency of FSConsG versus SVFG, revealing the extent to which our approach eliminates redundant constraints during graph construction. Total execution time measures the end-to-end duration required to complete the analysis for each benchmark program, serving as the primary indicator of computational efficiency. These complementary metrics enable us to systematically evaluate both the theoretical advantages of our constraint representation and its practical performance implications. By correlating reductions in constraint count with corresponding improvements in execution time, we establish a comprehensive assessment of \tool's effectiveness relative to VSFS.

\subsection{Research Questions}

Our experimental evaluation aims to address the following research questions:
\begin{enumerate}
    \item [RQ1] \textbf{Comparison of FSConsG and SVFG:} How does our FSConsG representation compare to the SVFG used in VSFS with respect to the number of constraints required for flow-sensitive pointer analysis?
    \item [RQ2] \textbf{Speedup of \tool:} To what extent does \tool improve analysis execution time compared to VSFS when integrated with state-of-the-art inclusion-based pointer analysis algorithms?
    \item [RQ3] \textbf{Memory Consumption of \tool:} What reduction in memory footprint does \tool achieve relative to VSFS during flow-sensitive pointer analysis?
\end{enumerate}

\subsection{Comparison of FSConsG and SVFG (RQ1)}

\begin{table*}[t]
  \centering
  \small
  \caption{Comparison of constraint numbers between SVFG (used in VSFS~\cite{barbar_object_2021}) and our proposed FSConsG.}
  \label{tab:numbers}
    \begin{tabular}{ l >{\raggedleft\arraybackslash}p{0.15\textwidth} rrr }
    \toprule
         \textbf{Project} & \textbf{SVFG}  & \textbf{FSConsG} & \textbf{Reduction} \\
          \midrule
    \texttt{lbm}       & 1,977      & 663       & 66.46\% \\
    \texttt{mcf}       & 12,713     & 8,004      & 37.04\% \\
    \texttt{namd}      & 279,903    & 174,211    & 37.76\% \\
    \texttt{deepsjeng} & 12,722     & 4,415      & 65.30\% \\
    \texttt{leela}     & 159,448    & 114,079    & 28.45\% \\
    \texttt{nab}       & 129,820    & 80,975     & 37.63\% \\
    \texttt{xz}        & 136,171    & 105,411    & 22.59\% \\
    \texttt{x264}      & 6,781,972   & 6,597,148   & 2.73\% \\
    \texttt{omnetpp}   & 185,970,609  & 151,739,481 & 18.41\% \\
    \texttt{povray}    & 12,021,871  & 7,798,511   & 35.21\% \\
    \texttt{cactus}    & 4,922,453   & 4,368,472   & 11.25\% \\
    \texttt{imagick}   & 26,823,216  & 14,985,392  & 44.13\% \\
    \midrule
    \textbf{\textit{Average}} &          &          & 33.91\% \\
    \bottomrule
    \vspace{-5mm}
    \end{tabular}%
\end{table*}%

\runinhead{Experimental Results.} Table~\ref{tab:numbers} presents a comparative analysis between FSConsG and SVFG in terms of constraint quantities across our benchmark programs. The data consistently demonstrates that FSConsG requires significantly fewer constraints than SVFG across all benchmarks, with an average reduction of 33.91\%. Notably, \texttt{lbm} exhibits the most substantial reduction at 66.46\%, where FSConsG eliminates the vast majority of constraints present in SVFG.
These findings empirically validate that FSConsG provides a substantially more compact representation than SVFG by systematically eliminating redundant constraints and consolidating necessary ones. This significant reduction in constraint count has profound implications for subsequent analysis stages, as the streamlined graph structure directly contributes to improved computational efficiency and enhanced scalability when performing flow-sensitive pointer analysis.

\runinhead{Results Analysis.} The substantial reduction in constraint count can be attributed to two key architectural decisions in our FSConsG design. First, FSConsG implements selective versioning that \emph{targets only address-taken variables}. Unlike SVFG, which introduces versions across a broader spectrum of program elements, FSConsG restricts versioning exclusively to address-taken variables—precisely those not already in SSA form that require flow-sensitive tracking. This targeted approach significantly reduces the proliferation of nodes that characterizes SVFG-based methods, yielding a more compact representation without sacrificing analytical precision.
Second, FSConsG employs a fundamentally different constraint encoding strategy by \emph{storing all constraint information exclusively on graph edges}. In contrast, SVFG-based approaches necessarily maintain constraints at both the edge and node levels to facilitate points-to set propagation. This dual-layer constraint representation in SVFG introduces substantial redundancy, whereas FSConsG's edge-centric design consolidates all constraint logic into a unified framework. The resulting streamlined architecture not only reduces the absolute constraint count but also simplifies subsequent propagation operations during analysis.

FSConsG's structural similarity to traditional constraint graphs enables the application of established graph simplification techniques. Because each edge in FSConsG directly encodes either a pointer operation or def-use relation, techniques such as cycle elimination and constraint folding can be applied systematically before each iteration of the flow-sensitive analysis. These optimizations further reduce the effective graph size and complexity during the analysis process, contributing to the observed efficiency gains.

\begin{answerBox}
\textbf{Answer to RQ1:} FSConsG results in significantly fewer constraints because it avoids the extensive node duplication inherent in SVFG-based approaches by confining versioning only to address-taken variables. Moreover, FSConsG exclusively stores constraint information on its edges, unlike SVFG, which must maintain constraints on both nodes and edges to support points-to set propagation. FSConsG enables effective graph simplification techniques such as cycle elimination and folding, further reducing the overall graph complexity.
\end{answerBox}

\subsection{Speedup of \tool (RQ2)}

\begin{table*}[t]
  \centering
  \small
  \caption{Comparison of execution time (in seconds) and speedup ratios between VSFS and \tool variants with wave propagation~\cite{pereira_wave_2009} (\tool-\wave) and SFR~\cite{chang_fast_2019} (\tool-\sfr).}
  \label{tab:rq2}
    \begin{tabular}{ l r r r }
        \toprule
        \textbf{Project}& \multicolumn{1}{c}{\textbf{VSFS}} 
        & \multicolumn{1}{c}{\textbf{\tool-Wave (Speedup)}} 
        & \multicolumn{1}{c}{\textbf{\tool-SFR (Speedup)}} \\
        \midrule
        \texttt{lbm}   & 0.012   & 0.007 (1.71\texttimes)  & 0.004 (3.00\texttimes) \\
        \texttt{mcf}   & 0.070   & 0.040 (1.75\texttimes)  & 0.018 (3.89\texttimes) \\
        \texttt{namd}  & 4.026   & 2.180 (1.85\texttimes)  & 1.163 (3.46\texttimes) \\
        \texttt{deepsjeng} & 0.138   & 0.064 (2.16\texttimes)  & 0.043 (3.21\texttimes) \\
        \texttt{leela} & 2.474   & 0.982 (2.52\texttimes)  & 0.375 (6.60\texttimes) \\
        \texttt{nab}   & 1.162   & 0.711 (1.63\texttimes)  & 0.240 (4.84\texttimes) \\
        \texttt{xz}    & 1.552   & 1.085 (1.43\texttimes)  & 0.258 (6.02\texttimes) \\
        \texttt{x264}  & 158.741 & 87.487 (1.81\texttimes)  & 16.617 (9.55\texttimes) \\
        \texttt{omnetpp} & 10067.5 & 6933.065 (1.45\texttimes)  & 501.629 (\textbf{20.07\texttimes}) \\
        \texttt{povray} & 302.244 & 111.240 (2.72\texttimes)  & 24.020 (12.58\texttimes) \\
        \texttt{ldecod} & 159.376 & 54.737 (\textbf{2.91\texttimes})  & 14.151 (11.26\texttimes) \\
        \texttt{imagick} & 136.985 & 116.377 (1.18\texttimes)  & 49.283 (2.78\texttimes) \\
        \midrule
        \textbf{\textit{Average}} &       & (1.93\texttimes)  & (7.27\texttimes) \\
    \bottomrule
    \end{tabular}%
    \vspace{-3mm}
\end{table*}

Table~\ref{tab:rq2} presents a comparative analysis of execution times between VSFS and \tool when integrated with two state-of-the-art constraint resolution algorithms: \wave Propagation (\wave)~\cite{pereira_wave_2009} and Stride-based Feld Representation (\sfr)~\cite{chang_fast_2019}. \tool exhibits significant speedup over VSFS in both scenarios, with an average 1.93\stimes improvement when using \wave and a remarkable 7.27\stimes speedup when employing \sfr. A key factor behind the superior performance of FSConsG over the SVFG-based approach lies in how each graph represents and propagates constraints. In FSConsG, constraint information is maintained exclusively on edges rather than being duplicated in both nodes and edges, as is the case with SVFG. This design choice reduces overhead and streamlines the process of points-to set propagation, particularly in iterative solvers such as \wave Propagation~\cite{pereira_wave_2009} and \sfr~\cite{chang_fast_2019}.  

\subsubsection{\tool-\wave}
The \wave Propagation algorithm~\cite{pereira_wave_2009} is an inclusion-based pointer analysis technique that enhances computational efficiency through three key mechanisms. First, it collapses strongly connected components to form an acyclic constraint graph, enabling a topologically ordered traversal. Second, it propagates pointer information incrementally by transmitting only the differences between current and previously propagated points-to sets, thereby minimizing redundant computations. Third, it dynamically updates the constraint graph by inserting new edges to represent complex pointer relationships, iterating until a fixed point is reached. 

\runinhead*{Experimental Results.}
Experimental results demonstrate that \tool-\wave consistently outperforms VSFS across all benchmarks, achieving an average speedup of 1.93\stimes. The performance improvements range from 1.18\stimes for \texttt{imagick} to 2.91\stimes for \texttt{ldecod}. Even for computationally lightweight benchmarks such as \texttt{lbm} and \texttt{mcf}, \tool-\wave exhibits measurable efficiency gains, reducing execution times from 0.012s to 0.007s and from 0.07s to 0.04s, respectively. For computationally intensive benchmarks, the performance advantage becomes more pronounced: analysis time for \texttt{povray} decreases from 302.244s to 111.24s (2.72\stimes speedup), while \texttt{omnetpp} improves from 10067.5s to 6933.065s (1.45\stimes speedup). These results demonstrate that the structural advantages of FSConsG translate directly into tangible performance benefits when leveraging the \wave Propagation algorithm.

\runinhead{Results Analysis.}
In the \wave Propagation algorithm, collapsing strongly connected components and performing topologically ordered traversals become more efficient when constraints are kept strictly on edges. Because FSConsG forgoes node-level constraints, fewer structures require maintenance or synchronization, thereby accelerating the propagation of pointer information and minimizing redundant work. As evidenced by the 1.93\stimes average speedup reported in Table~\ref{tab:rq2}, even smaller benchmarks (e.g., \texttt{lbm} and \texttt{mcf}) exhibit notable performance gains, while larger benchmarks like \texttt{omnetpp} and \texttt{povray} reap substantial time reductions.

\subsubsection{\tool-\sfr} 
The Stride-based Feld Representation (\sfr) algorithm~\cite{chang_fast_2019} enhances inclusion-based pointer analysis efficiency through strategic constraint graph decomposition. It identifies a minimal causality subgraph containing only nodes affected by recent updates, thereby localizing computation to relevant graph regions. \sfr systematically detects and collapses strongly connected components, transforming cyclic structures into an acyclic representation that facilitates topological ordering. This refined architecture significantly reduces computational redundancy during points-to set propagation by isolating updates to affected subgraphs, thus accelerating convergence while preserving the analysis's correctness, predictability, and scalability.

\runinhead{Experimental Results.}
The integration of \sfr with \tool yields substantial performance improvements, achieving an average speedup of 7.27\stimes over VSFS. Performance gains are consistent across the entire benchmark suite, ranging from 2.78\stimes for \texttt{imagick} to an exceptional 20.06\stimes for \texttt{omnetpp}. Even smaller benchmarks demonstrate marked improvements, with \texttt{lbm} and \texttt{mcf} achieving 3.00\stimes and 3.89\stimes speedups, respectively. Medium-sized programs like \texttt{x264} and \texttt{povray} show 9.55\stimes and 12.58\stimes gains. The impact on large-scale benchmarks is particularly noteworthy—\texttt{omnetpp}'s execution time decreases from 10067.5s to 501.629s, underscoring the exceptional scalability of our \sfr-based implementation when combined with the FSConsG representation.

\runinhead*{Results Analysis.} 
These results underscore the advantages of integrating FSConsG with the advanced \sfr optimization technique. By maintaining constraint information exclusively on graph edges and applying more aggressive simplifications, \tool-\sfr reduces redundant constraints and accelerates subsequent pointer analysis phases.
\sfr refines the constraint graph by identifying and collapsing strongly connected components into a minimal subgraph, focusing updates on the specific regions affected by new information. Because FSConsG already maintains a more compact, edge-centric graph, the decomposition and selective update steps in \sfr proceed with significantly less overhead compared to an SVFG-based structure. The result is an average 7.27\stimes speedup, including a remarkable 20.07\stimes acceleration for \texttt{omnetpp}, as shown in Table~\ref{tab:rq2}.  

\textbf{Performance Variation Across Benchmarks.}
The performance variation of CG-FSPTA across benchmarks is primarily influenced by
the program scale and the relative weight of the solving phase within the total
execution time. Our approach demonstrates the highest speedups (e.g., 20.07$\times$
for \texttt{omnetpp} with SFR) in large-scale benchmarks where the solving phase is
the primary bottleneck. In these cases, the streamlined FSConsG structure allows
advanced solvers to localize updates and collapse cycles much more efficiently than
on traditional SVFG structures.

Conversely, the reported speedup appears more modest on smaller benchmarks, such as
\texttt{lbm} and \texttt{deepsjeng}, despite achieving a constraint reduction of
over 65\% (Table~\ref{tab:numbers}). This is because the total execution time includes fixed
overheads for preprocessing (pre-analysis and FSConsG construction). For smaller
codebases, these setup costs represent a substantial portion of the end-to-end
duration, which masks the significant performance gains achieved during the actual
points-to solving phase. As the program scale increases, these fixed costs become
negligible, allowing the architectural efficiency of FSConsG to translate into
higher total speedup ratios. This observation is consistent across both the WAVE
and SFR integration results reported in Tables~\ref{tab:rq2}.

\begin{answerBox}
    \textbf{Answer to RQ2:} Our experimental results indicate that \tool substantially reduces execution time compared to VSFS when using inclusion-based pointer analysis optimizations. In particular, when combined with the \wave Propagation algorithm, \tool achieves an average speedup of 1.93\stimes. More strikingly, when using \sfr—which further refines the graph by collapsing strongly connected components and restricting updates to the minimal subgraph—\tool's average speedup rises to 7.27\stimes. In all cases, \tool preserves the same level of precision as VSFS, demonstrating that these speedups do not come at the expense of accuracy.
\end{answerBox}

\subsection{Memory Consumption of \tool (RQ3)}

\runinhead{Experimental Results.}
Table~\ref{tab:rq3} reveals that integrating FSConsG with the \wave Propagation algorithm (\tool-\wave) reduces memory consumption by an average of 23.99\% compared to VSFS. Notably, this improvement extends across both small and large benchmarks. For instance, \texttt{lbm} exhibits reductions exceeding 50\%, indicating that even simpler workloads can benefit substantially from \tool-\wave’s leaner constraint representation. Medium-scale applications such as \texttt{namd} and \texttt{deepsjeng} also show considerable savings, while larger, memory-intensive programs like \texttt{omnetpp} and \texttt{x264} realize more modest, yet still meaningful, reductions in peak memory usage.

\begin{table*}[t]
  \centering
  \small
  \caption{Comparison of memory consumption (in MB) and reduction ratios between VSFS and \tool variants with wave propagation~\cite{pereira_wave_2009} (\tool-Wave) and SFR~\cite{chang_fast_2019} (\tool-SFR).}
  \label{tab:rq3}
    \begin{tabular}{ l r r r }
        \toprule
        \textbf{Project}& \multicolumn{1}{c}{\textbf{VSFS}} 
        & \multicolumn{1}{c}{\textbf{\tool-Wave (Reduction)}} 
        & \multicolumn{1}{c}{\textbf{\tool-SFR (Reduction)}} \\
        \midrule
    \texttt{lbm}   & 8.81  & 4.20 (52.33\%) & 3.43 (61.07\%) \\
    \texttt{mcf}   & 25.53  & 18.00 (29.52\%) & 13.20 (48.32\%) \\
    \texttt{namd}  & 674.48  & 362.29 (46.29\%) & 333.91 (50.49\%) \\
    \texttt{deepsjeng} & 53.10  & 24.26 (54.31\%) & 22.97 (56.74\%) \\
    \texttt{leela} & 295.18  & 253.08 (14.26\%) & 186.12 (36.95\%) \\
    \texttt{nab}   & 217.22  & 181.54 (16.42\%) & 136.09 (37.35\%) \\
    \texttt{xz}    & 235.35  & 205.29 (12.77\%) & 160.91 (31.63\%) \\
    \texttt{x264}  & 10531.30  & 9565.82 (9.17\%) & 8643.88 (17.92\%) \\
    \texttt{omnetpp} & 35184.50  & 22733.10 (35.39\%) & 25601.00 (27.24\%) \\
    \texttt{povray} & 14419.50  & 13529.04 (6.18\%) & 12524.40 (13.14\%) \\
    \texttt{cactus} & 7343.16  & 6762.89 (7.90\%) & 6607.95 (10.01\%) \\
    \texttt{imagick} & 27003.70  & 26106.74 (3.32\%) & 25440.70 (5.79\%) \\
    \midrule
    \textbf{\textit{Average}} &      & 23.99\%      & 33.05\% \\
    \bottomrule
    \vspace{-5mm}
    \end{tabular}%
\end{table*}%

\begin{table*}[t]
  \centering
  \small
  \caption{Comparison of the number of resolved points-to sets of address-taken variables when using SVFG (as in VSFS~\cite{barbar_object_2021}) versus our proposed FSConsG.}
  \label{tab:pts}
    \begin{tabular}{ l >{\raggedleft\arraybackslash}p{0.15\textwidth} rrr }
    \toprule
         \textbf{Project} & \textbf{SVFG}  & \textbf{FSConsG} & \textbf{Reduction} \\
          \midrule
    \texttt{lbm}   & \multicolumn{1}{r}{159} & \multicolumn{1}{r}{140} & 11.95\% \\
    \texttt{mcf}   & \multicolumn{1}{r}{1,523} & \multicolumn{1}{r}{1,095} & 28.10\% \\
    \texttt{namd}  & \multicolumn{1}{r}{31,803} & \multicolumn{1}{r}{29,772} & 6.39\% \\
    \texttt{deepsjeng} & \multicolumn{1}{r}{638} & \multicolumn{1}{r}{577} & 9.56\% \\
    \texttt{leela} & \multicolumn{1}{r}{51,582} & \multicolumn{1}{r}{39,766} & 22.91\% \\
    \texttt{nab}   & \multicolumn{1}{r}{27,005} & \multicolumn{1}{r}{24,749} & 8.35\% \\
    \texttt{xz}    & \multicolumn{1}{r}{52,138} & \multicolumn{1}{r}{38,950} & 25.29\% \\
    \texttt{x264}  & \multicolumn{1}{r}{9,040,845} & \multicolumn{1}{r}{6,094,054} & \textbf{32.59\%} \\
    \texttt{omnetpp} & \multicolumn{1}{r}{51,825,737} & \multicolumn{1}{r}{47,848,627} & 7.67\% \\
    \texttt{povray} & \multicolumn{1}{r}{2,521,484} & \multicolumn{1}{r}{2,074,473} & 17.73\% \\
    \texttt{cactus} & \multicolumn{1}{r}{3,706,226} & \multicolumn{1}{r}{3,167,779} & 14.53\% \\
    \texttt{imagick} & \multicolumn{1}{r}{2,575,964} & \multicolumn{1}{r}{2,016,879} & 21.70\% \\
    \midrule
    \textbf{\textit{Average}} &       &       & 17.23\% \\
    \bottomrule
    \end{tabular}%
\end{table*}%

When combined with the Stride-based Feld Representation (\sfr) solver, \tool-\sfr demonstrates an even greater reduction in memory usage, achieving an average decrease of 33.05\% relative to VSFS. The gains are especially pronounced for complex or large-scale benchmarks. For example, \texttt{lbm} again shows over 60\% memory savings, and \texttt{omnetpp} and \texttt{x264} also see more substantial reductions under \sfr than under \wave. These results illustrate that \sfr’s strategy of localizing updates to a minimal subgraph—together with FSConsG’s inherently streamlined structure—can significantly lower the memory footprint of inclusion-based pointer analysis.

\runinhead{Results Analysis.}
The superior memory efficiency of \tool compared to VSFS stems primarily from fundamental architectural differences in constraint representation. While VSFS maintains points-to information for both input and output states of address-taken variables at store instructions~\cite{barbar_object_2021, sui_-demand_2016,hardekopf_flow-sensitive_2011}, FSConsG employs a more streamlined approach that stores points-to set information exclusively on nodes. Table~\ref{tab:pts} quantifies this difference in points-to sets of address-taken variables (while points-to sets for top-level variables remain identical across both approaches), revealing that FSConsG consistently requires fewer points-to sets across all benchmarks—averaging a 17.23\% reduction. This significant reduction in required points-to sets directly contributes to FSConsG's smaller memory footprint during analysis.

For \tool-\wave specifically, memory efficiency derives from two synergistic factors. First, the inherently compact FSConsG structure eliminates redundant constraint storage. Second, \wave's algorithm—which collapses strongly connected components and propagates only incremental pointer information—operates more efficiently on FSConsG's edge-centric representation. With fewer redundant constraints and simplified graph traversal paths, update operations consume substantially less memory during analysis. This architectural synergy between FSConsG and \wave's topological processing model explains the observed 23.99\% average reduction in memory consumption compared to VSFS.

\tool-\sfr achieves even greater memory efficiency by combining FSConsG's streamlined representation with \sfr's advanced graph processing techniques. The \sfr algorithm strategically decomposes the constraint graph and collapses cycles, confining points-to set propagation exclusively to graph regions affected by new pointer information. This localized processing approach, when applied to FSConsG's already-optimized structure, minimizes the memory footprint of in-process data structures during analysis. By restricting both the scope of active computation and the underlying graph representation, \tool-\sfr achieves a 33.05\% reduction in memory consumption while simultaneously delivering substantial performance improvements, demonstrating that FSConsG provides an ideal foundation for advanced inclusion-based pointer analysis algorithms.

\begin{answerBox}
    \textbf{Answer to RQ3:} \tool achieves significant memory consumption reductions compared to VSFS, averaging 23.99\% with the \wave algorithm and 33.05\% with \sfr. The improvements are primarily due to the FSConsG design, which confines versioning to address-taken variables and maintains constraint information exclusively on edges, thereby eliminating redundant node-level data structures.
\end{answerBox}

\section{Related Work}
\label{sec:related work}

This section examines prior research in two domains central to our work: Andersen-style pointer analysis, which formulates points-to relations as set-constraint problems, and flow-sensitive pointer analysis, which enhances precision by distinguishing points-to relationships at different program points. We discuss how our approach builds upon and differs from these established techniques.

\runinhead*{Andersen-Style Pointer Analysis.}
Andersen's style pointer analysis~\cite{andersen1994} is a widely-adopted and extensively-studied approach that formulates points-to relation resolution as a set-constraint or set-union problem by computing a dynamic transitive closure on the program's constraint graph. 
Numerous works have focused on improving this analysis at both algorithmic~\cite{hardekopf_ant_2007, andersen-precision, partial-cycle-elim, cla, pearce_efficient_2007, pereira_wave_2009, ovs, chang_fast_2019, andersen-complexity, liu2022pus,lei2024context} and data-structural~\cite{barbar2021compacting, barbar2021hash, doop, bdd, spark} levels to reduce the overhead of Andersen's style pointer analysis. These optimizations include employing efficient constraint solvers~\cite{pearce2004efficient, pereira_wave_2009, liu2022pus, chang_fast_2019}, simplifying constraint graphs through cycle elimination~\cite{partial-cycle-elim, hardekopf_ant_2007}, applying variable substitution~\cite{ovs, ander-equiv}, hash consing~\cite{barbar2021hash}, and points-to set compaction~\cite{barbar2021compacting}. Our FSConsG framework inherently benefits from these optimizations, as it employs a constraint graph architecture that naturally accommodates various Andersen's style pointer analyses.

\runinhead*{Flow-Sensitive Pointer Analysis.}
Flow-sensitive pointer analysis~\cite{zuo2021chianina, hardekopf_semi-sparse_2009, hardekopf_flow-sensitive_2011,barbar_object_2021,sui_sparse_2016,sui_value-flow-based_2018,choi1991automatic,hind1998assessing} enhances precision by distinguishing points-to relationships at different program control points. These approaches necessarily interact with program control flow during analysis. Given the computational complexity associated with analyzing multiple program points, researchers have proposed various efficiency-enhancing techniques.
Early research~\cite{choi1991automatic,hind1998assessing,strongupdate} focused on optimizing the interprocedural control flow graph (ICFG) through the elimination of analysis-irrelevant nodes. Subsequently, Hardekopf and Lin~\cite{hardekopf_semi-sparse_2009} introduced semi-sparse analysis, which applies sparse techniques to top-level pointers while retaining traditional analysis methods for address-taken objects.
Recent advances in sparse analysis~\cite{yao2024falcon,oh2012design,fink2008effective,hardekopf_flow-sensitive_2011,sui_value-flow-based_2018, sui_-demand_2016,barbar_object_2021} leverage def-use analysis to construct sparse value-flow graphs (SVFGs) and perform flow-sensitive analysis on the SVFG. These SVFGs encode def-use relationships between program variables while preserving program point information, with each SVFG node corresponding to a control flow graph node. In contrast, our approach eschews dependence on control flow or value flow graphs during the core flow-sensitive pointer analysis phase. Instead, we develop a novel constraint graph architecture that encodes requisite flow-sensitivity information offline as versioned address-taken variables, enabling the underlying pointer analysis algorithm to operate in a purely flow-insensitive Andersen's style without interaction with the control flow graph during analysis.

\section{Conclusion}
\label{Conclusion}
We present \tool, a technique performing flow-sensitive pointer analysis based on a flow-sensitive set-constraint graph FSConsG,
to address key limitations of traditional control-flow-based flow-sensitive pointer analysis. 
\tool integrates flow-insensitive and flow-sensitive pointer analyses to the same model, streamlining the analysis process by reducing graph complexity, memory usage, and execution time. These innovations enable significant improvements in computational efficiency and scalability while maintaining the analytical precision required for complex program analysis.
In particular, \tool achieves an average memory reduction of 33.05\% and accelerates flow-sensitive pointer analysis by 7.27\stimes compared to the state-of-art flow-sensitive pointer analysis method. Such results show the efficiency of \tool. Our study also suggests the potential for using set-constraint mind to solve flow-sensitive pointer analysis. This promising avenue represents an intriguing direction for future research in further enhancing analysis efficiency.


\bibliographystyle{cas-model2-names}
\bibliography{FSConsG}

\end{document}